\documentclass[twocolumn]{aastex701}

\usepackage{xcolor}
\def\mean#1{\left< #1 \right>}

\usepackage{amsmath} 

\shorttitle{Small-scale power spectrum and SKA}
\shortauthors{Zhu et al.}

\begin{document}

\title{Probing power spectrum enhancement at small scales with the SKA}

\correspondingauthor{Bin Yue, Zhiqi Huang}
\email{yuebin@nao.cas.cn, huangzhq25@mail.sysu.edu.cn}

\author[0009-0005-1646-6539]{Hourui Zhu}
\affiliation{State Key Laboratory of Radio Astronomy and Technology, National Astronomical Observatories, Chinese Academy of Sciences, 20A Datun
Road, Chaoyang District, Beijing 100101, China}
\affiliation{School of Astronomy and Space Science, University of Chinese Academy of Sciences, No.1 Yanqihu East Rd, Huairou District, Beijing 101408, China}
\email{zhuhr@bao.ac.cn} 

\author[0000-0002-7829-1181]{Bin Yue}
\affiliation{State Key Laboratory of Radio Astronomy and Technology, National Astronomical Observatories, Chinese Academy of Sciences, 20A Datun
Road, Chaoyang District, Beijing 100101, China}
\email{yuebin@nao.cas.cn}

\author[0000-0003-3224-4125]{Yidong Xu}
\affiliation{State Key Laboratory of Radio Astronomy and Technology, National Astronomical Observatories, Chinese Academy of Sciences, 20A Datun
Road, Chaoyang District, Beijing 100101, China}
\email{xuyd@nao.cas.cn}

\author[0000-0001-6475-8863]{Xuelei Chen}
\affiliation{State Key Laboratory of Radio Astronomy and Technology, National Astronomical Observatories, Chinese Academy of Sciences, 20A Datun
Road, Chaoyang District, Beijing 100101, China}
\affiliation{School of Astronomy and Space Science, University of Chinese Academy of Sciences, No.1 Yanqihu East Rd, Huairou District, Beijing 101408, China}
\email{xuelei@bao.ac.cn}

\author[0000-0002-1506-1063]{Zhiqi Huang}
\affiliation{School of Physics and Astronomy, Sun Yat-sen University, 2 Daxue Road, Tangjia, Zhuhai 519082, China}
\affiliation{CSST Science Center for the Guangdong-Hongkong-Macau Greater Bay Area, Sun Yat-sen University, Zhuhai 519082, China}
\email{huangzhq25@mail.sysu.edu.cn}

\begin{abstract}

The reionization process is driven by ionizing photons from dwarf galaxies in halos with virial temperature $T_{\rm vir} \gtrsim 10^4$ K, while minihalos whose $T_{\rm vir}\lesssim 10^4$ K consume ionizing photons and have negative contributions to reionization. Since ionizing sources and minihalos have different clustering characteristics, not only the reionization history, but also the morphology of the ionization field, is sensitive to the small-scale power spectrum. If the power spectrum at small scales is enhanced compared with the standard six-parameter $\Lambda$CDM model, then both the sources and sinks of ionizing photons would be boosted and the net impact depends on the competition between them. Therefore, the 21 cm signal that can probe the morphology of the ionization field will be a useful tool for detecting the small-scale power spectrum. Using the power spectrum   proposed by Cielo et al. (2025) (C25) as a demonstration, we investigate the influence of small-scale power spectrum enhancement on the ionization field and the 21 cm signal. We find that for the C25 model, even under the constraints of observed UV luminosity functions for high-$z$ galaxies and reionization history, the 21~cm power spectrum and the bubble size distribution could be still significantly different from the regular $\Lambda$CDM model. The upcoming SKA-low AA* telescope, and a further imaging telescope, have the potential to detect the small-scale power spectrum more deeply.  
\end{abstract}


\section{Introduction}

The matter power spectrum at the smallest scales remains an open question in cosmology. Primordial fluctuations are produced during the inflation stage, their evolution depends on the properties of dark matter particles. Although the Lambda cold dark matter ($\Lambda$CDM) model, with density fluctuations extending down to the damping scales of cold dark matter (CDM) particles (the mass scale of the CDM fluctuations damping is $\sim$an Earth mass, see \citealt{Diemand2005Natur,Wang2020Natur}),  has been used as the standard model in research (e.g. \citealt{planck18}), other models give rather different fluctuations levels at small scales. For example, in the Warm Dark Matter (WDM) model \citep{Bode2001ApJ,viel2005PhRvD,Smith2011PhRvD} or Fuzzy Dark Matter (FDM) model \citep{Jones2021ApJ,Gong2023ApJ}, the power spectrum starts to damp at scales much larger than the CDM damping scale. Therefore the formation of small dark matter halos are heavily suppressed. This has significant influence on the high-$z$ galaxy population \citep{Gong2023ApJ} and the reionization process \citep{Barkana2001ApJ,Yue2012ApJ}. Alternatively, there are also models with small-scale fluctuations enhanced significantly compared to the standard $\Lambda$CDM model (e.g. \citealt{yoshiurab2020PhRvD.101h3520Y,Inman2023PhRvD,Cielo2025,Nadler2025ApJ,minoda2023PhRvD.108l3542M}). The enhancement of the small-scale power spectrum could be an interpretation of the observed excess of galaxies at high redshifts \citep{hirano2024,tkachev2024}, or result in earlier formation of first stars and onset of the Cosmic Dawn.

The abundance of the high-$z$ galaxy population, i.e. the UV luminosity functions (UV LFs), can be a probe of small-scale power spectrum and dark matter properties \citep{Dayal2017ApJ,Rudakovskyi2021MNRAS,yoshiura2020PhRvD.102h3515Y,Dayal2024MNRAS,Liu2025arXiv}. The James Webb Space Telescope (JWST) has detected many faint high-$z$ galaxies (e.g. \citealt{naidu2022,finkelstein2022,castellano2022,bradley2023,atek2023}), however, they are actually classified as the rare and brightest ones among the high-$z$ galaxy population. Their host dark matter halos are $\gtrsim 10^{10}~M_\odot$ \citep{Endsley2020MNRAS}. Observing power spectrum at smaller scales is rather challenging, because small-scale structures generally host very faint galaxies or even do not host any luminous objects. The Ly$\alpha$ forest is considered as the most powerful tool for probing subgalactic-scale power spectrum, and it has detect structures down the $k\sim 1-10$ Mpc$^{-1}$ \citep{viel2005PhRvD,Chabanier2019MNRAS}, WDM particles mass up to $\sim 3-5$ keV \citep{viel2005PhRvD,Viel2013PhRvD,Irvic2017PhRvD,Villasenor2023PhRvD}, or FDM mass up to $\sim 10^{-21}-10^{-23}$ eV \citep{Lazare2024PhRvD,Sipple2025MNRAS}. But the power spectrum at $k \gtrsim 10$ Mpc$^{-1}$ is still not known. In the future, with the Square Kilometre Array (SKA) telescope, the 21 cm forest could be a powerful probe for either the damping \citep{shimabukuro2014PhRvD..90h3003S,Shimabukuro2020PhRvDa,shimabukuro2025PhRvD.112f3557S} or the enhancement \citep{Shimabukuro2020PhRvDb} of the small structures before the end of reionization. 

Reionization is the last global phase transition in the history of the Universe, during which the intergalactic medium (IGM) evolves from a predominantly neutral state to a fully ionized plasma, driven by ionizing photons from the first stars, galaxies and black holes \citep{barkana2001}. 
In the standard $\Lambda$CDM scenario, ionizing photons are  primarily provided by star-forming galaxies in dark matter halos with virial temperature  $T_{\mathrm{vir}} \gtrsim 10^4\,\mathrm{K}$, corresponding to halo masses $M_\mathrm{h} \gtrsim 10^8\,M_\odot$ in the epoch of reionization (EoR) \citep{Witstok2025Natur,Wu2024OJAp,Choustikov2025MNRAS}.
However, a population of lower-mass halos with $T_{\rm vir} \lesssim 10^4$ K, named minihalos, also plays a critical role in reionization process. While a fraction of minihalos with  $10^3~{\rm K} \lesssim T_{\rm vir} \lesssim 10^4~{\rm K}$ may host Pop III stars \citep{klessen2023}, most of them have no chance to host any stars before virial temperature exceeding $10^4$ K and Hydrogen atomic-cooling being efficient. 
Their primary impact on reionization is as potent sinks of ionizing photons, owing to recombinations in their dense, self-shielded gas \citep{barkana2002,shapiro2004,iliev2005,ciardi2006,yb2009,my2020,park2016ApJ...831...86P}. This sink behavior arises from two key mechanisms: self-shielding and enhanced recombination. Self-shielding occurs due to high gas densities that shield interiors from ionizing radiation, trapping ionization fronts \citep{shapiro2004,iliev2005}; while enhanced recombination results from elevated densities that increase recombination rates, leading to a large amount of ionizing photons consumption \citep{chan2024,park2021ApJ...908...96P}. If the power spectrum at small scales is enhanced compared to the regular $\Lambda$CDM model, then either the production of ionizing photons  by faint galaxies, or the consumption of ionizing photons by minihalos, or both of them, would be boosted. The net impact on reionization depends on the competition between the two effects, and the large-scale morphology of the ionization field would change. Observing the large-scale morphology of the ionization field provides potential possibility to detect power spectrum at small scales.

In the Dark Ages, the 21 cm emission line from neutral Hydrogen traces the density field very well, it can directly detect small-scale structures provide that the telescope has high angular resolution \citep{Loeb2004PhRvL,Cole2021MNRAS,park2025NatAs...9.1723P}. More promisingly, even without directly resolving structures at these scales, the large-scale power spectrum of the 21 cm signal also tells us the information of them  in the Dark Ages and the Cosmic Dawn \citep{Ali-Haimoud2014PhRvD,Munoz2020PhRvD,zhangxin2023,Sikder2025arXiv}. This is also the case for the EoR, where the 21 cm signal measures the large-scale morphology of the ionization field (e.g. \citealt{furlanetto2006, pritchard2012}) and provide us with information of faint galaxies and minihalos at small scales.

In this paper, we investigate the galaxy UV LFs at the EoR, the CMB scattering optical depth, the reionization history, the 21 cm power spectrum and bubble size distribution (BSD) in a small-scale enhanced power spectrum model proposed by \citet{Cielo2025} (C25 hereafter). We compare these results with the regular $\Lambda$CDM model and estimate the detectability of their differences for the upcoming SKA-low AA* and an assumed further telescope that is able to do tomographic imaging observations.
This paper is organized as follows: in Section~\ref{sec:method}, we introduce our methods for constructing the UV LFs of galaxies during the EoR from the power spectrum model, and for building ionization and 21 cm fields using the semi-numerical algorithm. We take into account the boost in both ionizing photons production by faint galaxies, and consumption by minihalos. In Section~\ref{sec:result}, we present our results on the reionization history, the 21 cm power spectrum, and the BSD. We also analyze the detectability of the small-scale power spectrum effects through these quantities. We give the summary and discussion in Section~\ref{sec:summary}. Throughout this paper, we adopt cosmological parameters from Planck18 release, say ($\Omega_{\rm m}$, $\Omega_{\rm b}$, $\Omega_\Lambda$, $h$, $\sigma_8$, $n_{\rm s}$) $=$ ($0.315$, $0.0493$, $0.685$, $0.674$, $0.811$, $0.965$) \citep{planck18}. 

\section{Methods}
\label{sec:method}

In this section, we introduce the model proposed in \citet{Cielo2025}, for which the small-scale density fluctuations are enhanced compared to the regular $\Lambda$CDM model. We then  calculate the UV LFs  and ionization fields in such a model, and the corresponding 21 cm signal fields.

\subsection{The C25 model and halo mass function}

\citet{Cielo2025} proposed a model for which the primordial scalar power spectrum is enhanced on small scales by sudden phase transitions during inflation. Considering de Sitter-invariant initial state for the inflaton field in $\alpha$ vacua~\citep{Allen1985} - which can be excited by preceding transitions - \citet{Cielo2025} find a $\sim k^6$ growth in a large volume of parameter space. 

For de Sitter-invariant initial state in $\alpha$ vacua, the normalization and phase difference of Bogoliubov coefficients are parameterized by two real parameters $\alpha$ and $\beta$, respectively. In the simplified case where the phase difference vanishes ($\beta=0$), and when the phase transition is modeled as a sudden change in the slope of the inflaton potential, the primordial scalar power per e-fold is given by 

\begin{align}
\Delta^2_{\rm C25}(k) = & \frac{9H^6}{4\pi^2 A_1^2 A_2^2} \Bigg\{ e^{-2\alpha} \bigg[ A_1 \nonumber \\
& + 3(A_1 - A_2)\frac{(\kappa^2 - 1)\sin(2\kappa) + 2\kappa\cos(2\kappa)}{2\kappa^3} \bigg]^2 \nonumber \\
& + e^{2\alpha} \left[ 3(A_1 - A_2)\kappa \, {\rm j}_1^2(\kappa) \right]^2 \Bigg\},
\label{eq:P_R}
\end{align}
where $A_1$ and $A_2$ are respectively the pre- and post-transition potential slopes, and $\mathrm{j}_1$ is the spherical Bessel function of order $1$. The dimensionless wavenumber $\kappa$ is defined as $k / k_{\text{trans}}$, where $k_{\rm trans}$ is the comoving horizon scale at the phase transition. The term $9H^6/(4\pi^3A_1^2A_2^2)$ in the right-hand side of Eq.~(\ref{eq:P_R}) is a normalization factor in perfect de Sitter background. In more realistic slow-roll scenarios, this normalization factor slowly varies and can be approximated as a power-law function with spectral index $n_s-1$, whose amplitude is much smaller than $1$.

For large excitation ($\alpha \gg 0$) and a strong transition ($A_1 \gg A_2$), the spectrum can grow as $\Delta^2_{\rm C25}(k) \propto k^6$ before reaching a peak, significantly exceeding the $k^4$ scaling in  the case of no excitation ($\alpha=0$, Bunch-Davies vacuum).  In the absence of phase transition ($A_1=A_2$) and with the incoming state being Bunch-Davies vacuum, the power spectrum reduces into the regular nearly scale-invariant model. To conduct a concrete study of this model, throughout this paper we take representative parameter values $A_2/A_1=0.1$ and $\alpha=2$ which give rise to the $k^6$ growth, and only vary $k_{\rm trans}$.

\begin{figure}

\centering

\includegraphics[width=0.45\textwidth]{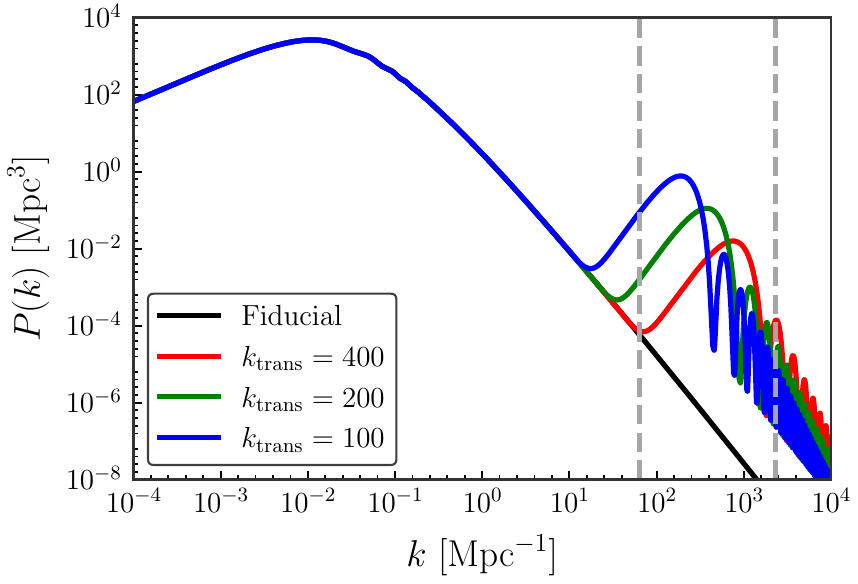}
\caption{
The power spectrum at $z=6$ for the fiducial model (regular $\Lambda$CDM model) and the C25 model with different $k_{\rm trans}$ values.
The vertical dashed lines refer to scales of different typical masses: the left line corresponds to the virial mass at virial temperature $T_\mathrm{vir} = 10^4 \, \mathrm{K}$, set as the lower mass limit for reionization sources and upper limit of minihalos; the right line corresponds to the Jeans mass, representing the minimal mass of minihalos hosting gas.
}
\label{fig:powspec}
\end{figure}

The linear matter power spectrum in this C25 model can be obtained by rescaling the primordial scalar power spectrum,
\begin{equation}
P_{\rm C25}(k,z)=P(k,z) \times \frac{\Delta^2_{\rm C25
}(k)}{\Delta^2(k)},
\end{equation}
where $P(k,z)$ and $\Delta^2(k)$ are matter power spectrum and dimensionless primordial power spectrum in the regular $\Lambda$CDM model respectively. We calculate $P(k,z)$ and $\Delta^2(k)$ from {\tt CAMB} \citep{lewis2000} for $k < 100\,\mathrm{Mpc}^{-1}$, and extrapolate to higher $k$ using the fitting formula BBKS \citep{bbks}. 
In Figure \ref{fig:powspec} we plot the matter power spectrum at $z=6$ for the regular $\Lambda$CDM model, used as the fiducial model in this paper, and for the C25 models with different $k_{\rm trans}$ values. In the C25 model, the matter power spectrum is significantly enhanced at small scales. Toward small scales, it starts to obviously deviate from the fiducial model at $k_{\rm start}\sim 0.15k_{\rm trans}$; reaches a peak and forms a ``bump'' at $k_{\rm bump}\sim 1.9k_{\rm trans}$; and then extends to much smaller scales.

With the matter power spectrum, we then calculate the halo mass function from the Sheth \& Tormen formalism \citep{st1999}, 
\begin{equation}
    \frac{\mathrm{d}n}{\mathrm{d}M_{\rm h}}(M_{\rm h}, z) = \rho_{\rm m} \frac{1}{M_{\rm h}} \frac{1}{\sigma(M_{\rm h})}  \left |\frac{\mathrm{d}\sigma(M_{\rm h})}{\mathrm{d}M_{\rm h}}  \right|f_{\rm ST}(\sigma, z)
    \label{eq:ps_mass_function}
\end{equation}
for which 
\begin{equation}
    f_{\rm ST}(\sigma,z) = A \sqrt{\frac{2a}{\pi}} \left[ 1 + \left( \frac{\sigma^2}{a\delta_c^2} \right)^p \right] \frac{\delta_c}{\sigma} \exp\left( -\frac{a\delta_c^2}{2\sigma^2} \right),
    \label{eq:sheth_tormen_f}
\end{equation}
and 
\begin{equation}
    \sigma^2(M_{\rm h}, z) = \frac{1}{2\pi^2} \int_0^\infty k^2 P(k) W^2(k, M_{\rm h}) \, ~\mathrm{d}k.
    \label{eq:mass_variance}
\end{equation}
In above equations, $\delta_c = 1.686 / D(z)$ is the halo collapse threshold linearly extrapolated to $z=0$. Parameters $A = 0.353$, $a = 0.73$, and $p = 0.175$ are fitted from the $N$-body simulations   \citep{jenkins2001}.

\begin{figure}
\centering
\includegraphics[width=0.45\textwidth]{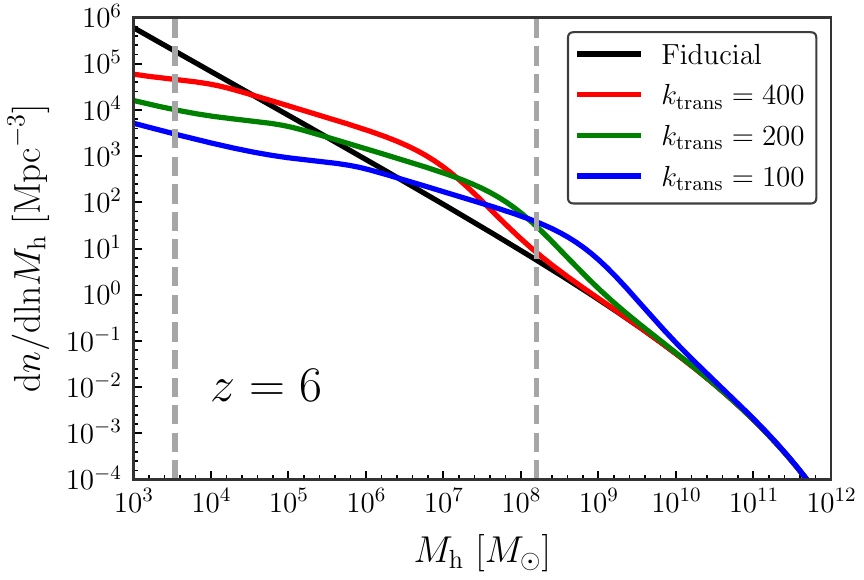}\\
\includegraphics[width=0.45\textwidth]{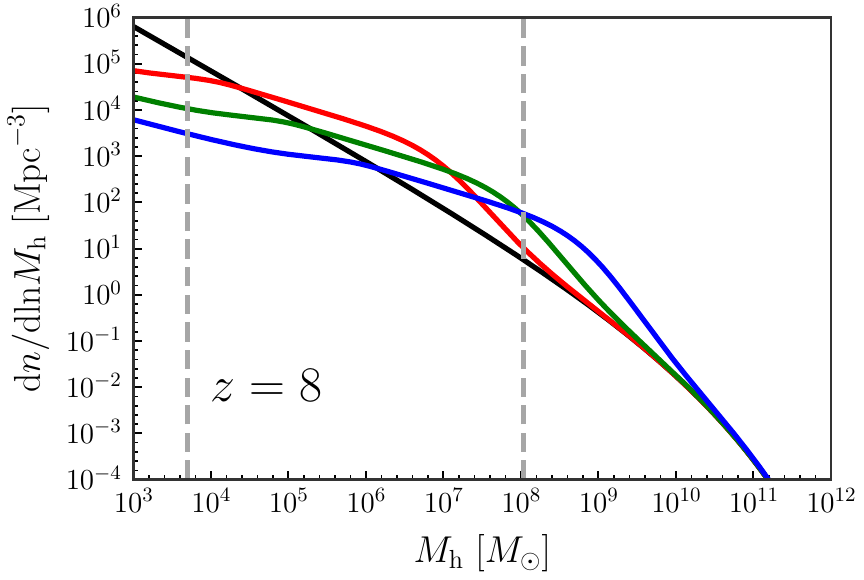}\\
\includegraphics[width=0.45\textwidth]{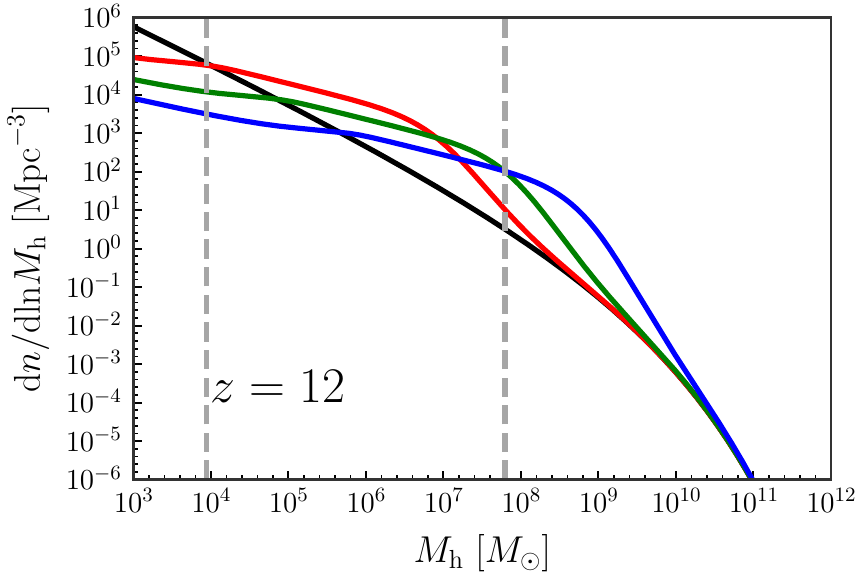}\\
\caption{Halo mass functions for fiducial model and C25 model with different $k_{\rm trans}$ values at the EoR. In each panel, the left vertical line refers to the Jeans mass, which is the minimal mass of minihalos able to retain gas; the right vertical line corresponds to virial mass of $T_\mathrm{vir} = 10^4 \, \mathrm{K}$, set as the lower mass limit for reionization sources and upper limit of minihalos.
}
\label{fig:hmf}
\end{figure}

In Figure \ref{fig:hmf} we show the halo mass functions in the fiducial  model and the C25 models with different $k_{\rm trans}$ values. 

Compared with the fiducial model, in the C25 model the number of halos is significantly boosted in a certain mass range, but heavily depressed at the mass scale much smaller than this range.  From higher mass to lower mass, the mass function starts to be obviously boosted at the  mass scale corresponding to $k_{\rm start}$,
\begin{align}
M_{\rm h}^{\rm start}&\sim\rho_{\rm m}\frac{4\pi}{3} \left( \frac{2\pi}{k_{\rm start}} \right)^3 \nonumber \\
&\approx \rho_{\rm m}\frac{4\pi}{3} \left( \frac{2\pi}{0.15k_{\rm trans}} \right)^3 \nonumber \\
&\approx300\rho_{\rm m}\frac{4\pi}{3} \left( \frac{2\pi}{k_{\rm trans}} \right)^3.
\end{align}
The boost keeps increasing until a mass point $M_{\rm h}^{\rm bend}$, then the halo mass function in the C25 model starts bend to the fiducial model. Below $\sim M_{\rm h}^{\rm bend}$, in the C25 model the halo number density grows more slowly than in the fiducial model, and finally becomes lower than the fiducial model.

From a mathematical perspective, according to Eqs. (\ref{eq:ps_mass_function} \& \ref{eq:sheth_tormen_f}), the halo number density is primarily governed by the term $|\mathrm{d}\sigma/\mathrm{d}M_\mathrm{h}|/\sigma^{2(1-p)}$. In the fiducial model, this term gradually grows with decreasing $M_{\rm h}$. 
In the C25 model, when $M_{\rm h}$ is approaching $M^{\rm bend}_{\rm h}$ from the higher mass side, both $|\mathrm{d}\sigma/\mathrm{d}M_\mathrm{h}|$ and $\sigma^{2(1-p)}$ rise to higher than the fiducial model. At the beginning,  $|\mathrm{d}\sigma/\mathrm{d}M_\mathrm{h}|$ increases faster than $\sigma^{2(1-p)}$, so the halo number density grows more rapidly than the fiducial model. 
However, when $M_{\rm h}$ drops to  $\lesssim M_{\rm h}^{\rm bend}$, $\sigma^{2(1-p)}$ exhibits a steep rise. So the growth rate of the ratio $|\mathrm{d}\sigma/\mathrm{d}M_\mathrm{h}|/\sigma^{2(1-p)}$ slows down to below the fiducial model.  

To derive the $M_{\rm h}^{\rm bend}$, we note that the behavior of $\sigma^{2(1-p)}$ is similar to $\sigma^2$. From the Eq. (\ref{eq:mass_variance}), the contribution of the power spectrum to the $\sigma^2$ is controlled by the window function $W^2(k,M_{\rm h})$. Only when $k\lesssim 0.5\times 2\pi/R_{M_{\rm h}}$, where $R_{M_{\rm h}}$ is the radius scale of the mass $M_{\rm h}$, the modes have sufficient contribution in the $\sigma^2$. Therefore the $\sigma^2$ starts to increases suddenly only when the modes with enhanced power spectrum ($k \gtrsim k_{\rm start}$) is covered by the window function. So the bend point must satisfy 
\begin{align}
k_{\rm start}\sim 0.15 k_{\rm trans} \lesssim 0.5 \frac{2\pi}{R_{M^{\rm bend}_{\rm h}}}.
\end{align}
Therefore 
\begin{align}
M_{\rm h}^{\rm bend}&=\rho_{\rm m}\frac{4\pi}{3} R^3_{M^{\rm bend}_{\rm h}} \nonumber \\
&\sim \rho_{\rm m}\frac{4\pi}{3} \left(\frac{2\pi}{0.3k_{\rm trans}}\right)^3 \nonumber \\
&\approx 40\rho_{\rm m}\frac{4\pi}{3} \left(\frac{2\pi}{k_{\rm trans}}\right)^3.
\end{align}

Physically speaking, in the C25 model, although the matter power spectrum enhancement boosts the collapse probability at all mass scales, it also accelerates the hierarchical merger process. Smaller halos merger into larger ones more frequently and even induces a deficit of smaller halos in the C25 model compared with the fiducial model. Actually, similar phenomenon is also observed in the fiducial model itself. Although the matter power spectrum increases monotonously with decreasing redshift, at $z\sim 0$, the number density of halos below $\sim 10^7~M_\odot$ is even smaller than higher redshifts, because of the hierarchical merger.

Explicitly, for $k_{\rm trans}=100$ Mpc$^{-1}$, in the EoR halos with mass $10^{6}M_{\odot} \lesssim M_{\rm h} \lesssim 10^{10}M_{\odot}$ are enhanced, while halos with mass $M_{\rm h}\lesssim 10^{6}M_{\odot}$ are depressed. In this case the faint-end of the UV LFs will be enhanced, and quite likely that reionization will be promoted.
For $k_{\rm trans}=200$ Mpc$^{-1}$, in the mass range $10^{5}M_{\odot} \lesssim M_{\rm h} \lesssim 10^{9}M_{\odot}$ halos are enhanced and, below $10^{5}M_{\odot}$ halos are depressed. The UV LFs at the fain-end will be enhanced but the reionization is more complicated. We will see later on that in this case, reionization is promoted at the beginning, however be delayed at the late stage of the EoR. For $k_{\rm trans}=400$ Mpc$^{-1}$, halos in $10^{5}M_{\odot} \lesssim M_{\rm h} \lesssim 10^{8}M_{\odot}$ are enhanced. This mass range is below the minimum halo mass for galaxy formation and entirely confined to minihalo mass. Then the UV LFs are not influenced but the reionization process will be delayed, as the minihalos would consume ionizing photons.

\subsection{The UV LFs}

Similar to \citet{zm2025}, we derive the UV LFs from halo mass functions, based on SFR--$M_{\rm h}$ and $L_{\rm UV}$--SFR relations, where $L_{\mathrm{UV}}$ is the UV luminosity defined at $1600\,\text{\AA}$, and SFR is the star formation rate.
The SFR--$M_\mathrm{h}$ relation is expressed as:
\begin{equation}
\mathrm{SFR}(M_\mathrm{h},z) = f_{*}(M_\mathrm{h},z) \frac{\Omega_b}{\Omega_m} \dot{M}_\mathrm{h},
\end{equation}
where the dark matter halo accretion rate \citep{mcbride2009}
\begin{align}
\dot{M}_\mathrm{h}(M_\mathrm{h},z)= &24.1\left(\frac{M_\mathrm{h}}{10^{12}\,M_\odot}\right)^{1.094} \times \nonumber \\
&   (1+1.75z)\sqrt{\Omega_m(1+z)^3+\Omega_\Lambda}\;[M_\odot\,\mathrm{yr}^{-1}]
\label{eq:mdot}
\end{align}
is fitted from N-body simulations,
and the star formation efficiency, $f_*$, follows the double power law formalism (e.g. \citealt{tacchella2018}):
\begin{equation}
f_{*}(M_\mathrm{h}) = 2f_0 \left[ \left( \frac{M_\mathrm{h}}{M_{\rm c}} \right)^{-\gamma_{\mathrm{lo}}} + \left( \frac{M_\mathrm{h}}{M_{\rm c}} \right)^{\gamma_{\mathrm{hi}}} \right]^{-1}.
\label{sfe}
\end{equation}
Here  $f_0=0.12$ is the star formation efficiency at the characteristic mass $M_{\rm c}=10^{11.7}~M_\odot$, $\gamma_{\mathrm{lo}}=0.66$ and $\gamma_{\mathrm{hi}}=0.65$ are low-mass and high-mass slopes respectively.

For a halo with mass $M_{\rm h}$ at redshift $z$, we assume that the log of UV luminosity has a Gaussian distribution around a central value, 
\begin{equation}
P_{\log L_{\mathrm{UV}}} = \frac{1}{\sqrt{2\pi}\sigma_{\mathrm{UV}}} \exp\left( -\frac{(\log L_{\mathrm{UV}} - \log \hat{L}_{\mathrm{UV}})^2}{2\sigma_{\mathrm{UV}}^2} \right),  
\label{eq:P_logL}
\end{equation}
where we set the scatter $\sigma_\mathrm{UV}=0.30$ and   
$\hat{L}_{\mathrm{UV}}$ is given by
\begin{equation}
\hat{L}_{\mathrm{UV}} = \mathrm{SFR} / \mathcal{K}_{\mathrm{UV}},  
\label{eq:lcenter}
\end{equation} 
and $\mathcal{K}_{\mathrm{UV}} = 1.17 \times 10^{-28}~M_\odot~\mathrm{yr}^{-1}/(\mathrm{erg}~\mathrm{s}^{-1}~\mathrm{Hz}^{-1})$ \citep{zm2025}.

The UV LF is obtained by integrating over halo mass: 
\begin{equation}
\frac{\mathrm{d}n}{\mathrm{d}\log L_{\mathrm{UV}}} = \int_{M_{\rm min}}^\infty \mathrm{d}M_\mathrm{h} \,f_{\mathrm{duty}}(M_\mathrm{h}, z) \frac{\mathrm{d}n}{\mathrm{d}M_\mathrm{h}} P_{\log L_{\mathrm{UV}}}
\label{eq:11}
\end{equation}  
where $M_{\rm min}$ is the minimum mass of halos allowing galaxy formation, and we adopt the virial mass corresponding to virial temperature $10^4$ K \citep{barkana2001}.  
\begin{equation}
    f_\mathrm{duty}(M_\mathrm{h},z)=\exp \left(-\frac{ M_\mathrm{turn}(z)}{M_\mathrm{h}}\right),
    \label{eq:fdy}
\end{equation}
is the duty cycle \citep{park2019}, i.e., only a fraction $f_{\rm duty}$ of halos host active star formation.
The turnover threshold mass \citep{cruz2025} 
\begin{equation}
M_\mathrm{turn}(z)=3.3\times10^7 \left[\frac{(1+z)}{21}\right]^{-3/2} ~M_\odot.
\end{equation}

We further add the dust effects via
\begin{equation}
    M_{\mathrm{UV}} = M_{\mathrm{UV}}^{\mathrm{obs}} - A_{\mathrm{UV}},
\end{equation}
where the dust attenuation
\begin{equation}
    A_{\mathrm{UV}} = C_1 + C_0 \beta, \, (A_\mathrm{UV}\ge0).
\end{equation}
We adopt the coefficients $C_0 = 2.10$ and $C_1 = 4.85$ \citep{koprowski2018}, and the UV spectral slope $\beta$ is taken from \citet{vogelsberger2019}.

\begin{figure*}[ht!]
\includegraphics[width=0.47\textwidth]{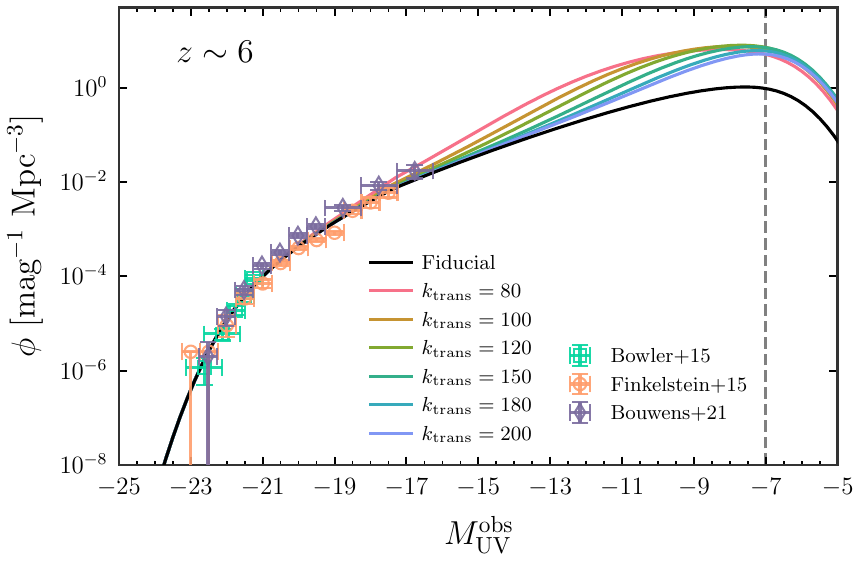}
\hfill
\includegraphics[width=0.47\textwidth]{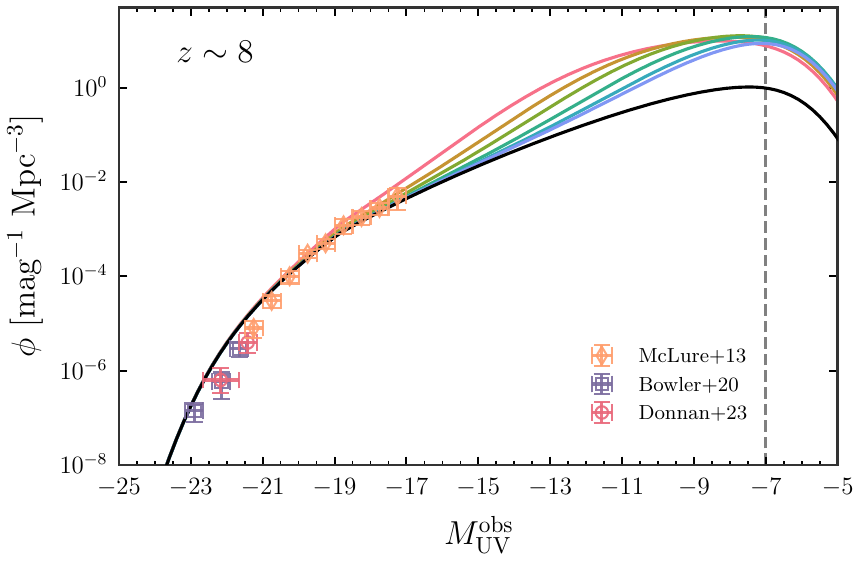}
\includegraphics[width=0.47\textwidth]{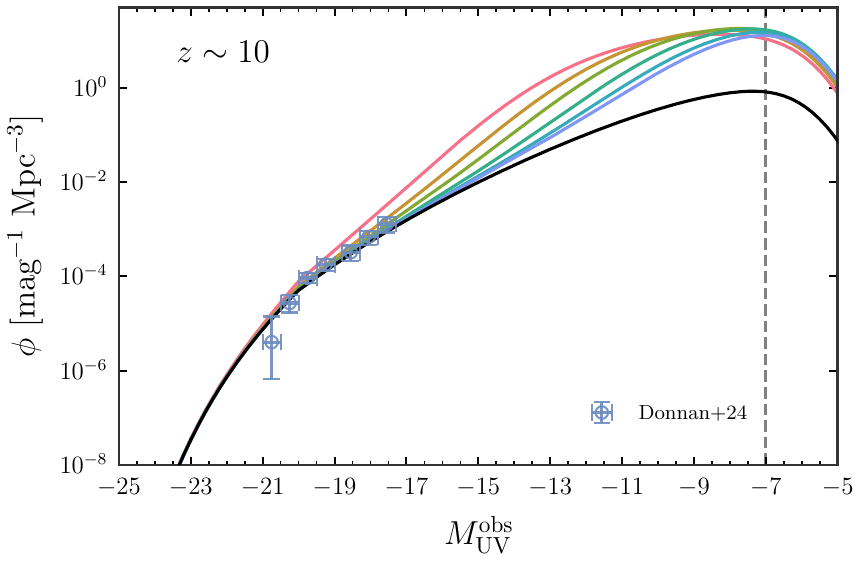}
\hfill
\includegraphics[width=0.47\textwidth]{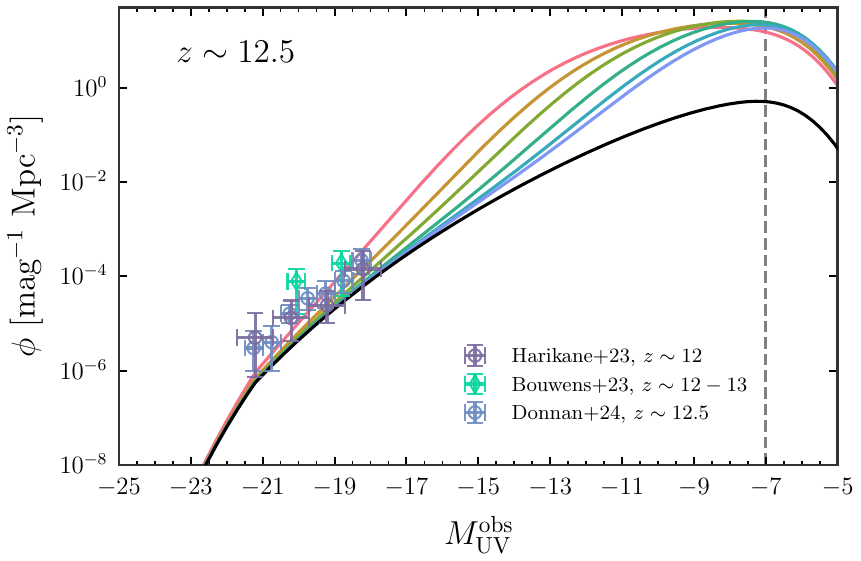}
\hfill
\caption{
The predicted UV LFs in fiducial model and in C25 models with different values of $k_{\mathrm{trans}}$, compared with observations \citep{mclure2013,finkelstein2015,bowler2020,bouwens2021,harikane2023,bouwen2023,donnan2023,donnan2024}. The vertical dashed line in each panel indicates the central absolute UV magnitude corresponding to the minimum halo mass required for star formation, $\hat{M}_{\rm UV}\sim -7$.}

\label{fig:uvlf_grid}
\end{figure*}

In Figure~\ref{fig:uvlf_grid} we show the predicted UV LFs for the fiducial model and the C25 models with different $k_{\rm trans}$ values, compared with the observations of HST and JWST \citep{mclure2013,finkelstein2015,bowler2020,bouwens2021,harikane2023,bouwen2023,donnan2023,donnan2024}. The $L_{\rm UV}$ is converted to absolute UV magnitude.

In the fiducial model, the predicted UV LFs agree with the observations very well, and as long as $k_{\mathrm{trans}} \gtrsim 100$ Mpc$^{-1}$,  UV LFs in C25 models are also consistent with observations. Interestingly, at the redshifts $\gtrsim 12$, the UV LFs in C25 models agree with observations better than the fiducial model, providing a potential solution for the excess of high redshift galaxy population in JWST observations (e.g. \citealt{Adamo2025NatAs,harikane2023}).

In our model, the typical absolute UV magnitude of the minimum halo mass allowing star formation, $\hat{M}_{\rm UV}$, is $\sim -7$, where $\hat{M}_{\rm UV}$ is the absolute magnitude of the central UV luminosity $\hat{L}_{\rm UV}$ in Eq. (\ref{eq:P_logL}), and this value just changes slightly in the redshift  range considered in this paper.
However, because we assume real magnitudes $M_{\rm UV}$ follow Gaussian distribution around this central value, the faintest magnitude can extend to far below this central value, just the probability is very small. At the faint-end, the UV LFs in Figure \ref{fig:uvlf_grid} start to deviate from the power-law faint-end in the popular Schechter formula (e.g. \citealt{bouwens2015ApJ...803...34B,finkelstein2015})  at $M_{\rm UV}$ from $\sim -13$ to $-8$. This is broadly consistent with predictions in semi-analytical models (e.g. \citealt{yung2019MNRAS.483.2983Y}) and simulations (e.g. \citealt{Ocvirk2020MNRAS.496.4087O}). We note that in many simulations the UV LFs drop dramatically below $M_{\rm UV}\sim -10$ (e.g \citealt{gnedin2016ApJ...825L..17G}), however our UV LFs still gradually grow or at least keeps flat until $M_{\rm UV}\sim -7$, probably because we do not model the complicated feedback mechanisms. Currently, the deep observations with strong gravitational lensing probe the UV LFs down to magnitudes as faint as $M_{\rm UV} \sim -13$ \citep[e.g.,][]{ Atek2018MNRAS.479.5184A, ishigaki2018ApJ...854...73I, bouwens2022ApJ...940...55B}. Other works aiming to reconstruct the EoR UV LFs from the observations of local dwarf galaxies extend the faintest magnitudes down to $\sim -5$ (e.g. \citealt{weisz2014ApJ...794L...3W}). But none of those works have yet confirmed the deviation from the power-law, as the uncertainties are still large \citep{Livermore2017ApJ,bouwens2017ApJ...843..129B,Yue2018ApJ...868..115Y}.
The shape of the UV LFs faint-end reflects the feedback mechanisms that are more efficient in smaller halos \citep{yue2016MNRAS.463.1968Y}. It can be described by involving more sophisticated models of the duty-cycle, star formation mode and efficiency in our model.

\subsection{The reionization process and the 21 cm signal}

To compute the 21 cm brightness temperature field in the EoR, we first generate the ionization field using a semi-numerical excursion-set approach \citep{fzh04,mesinger2011}. 
In the excursion-set framework, a spherical region with radius $R$ is ionized as long as the cumulative number of ionizing photons escaped into the IGM is larger than the consumptions by both the IGM and minihalos,
\begin{align}
N_{\rm ion}( z|R,\delta_R)\ge & (1+\bar{n}_{\rm rec})\,N_{\mathrm{H}}(\delta_R) (1-f_{\rm coll}({z|>M_{\rm J}}))\nonumber \\  
& +N_{\xi}(z| R,\delta_R),
\label{eq:excursion minihalo}
\end{align}
where $N_{\rm ion}( z| R,\delta_R)$ denotes the cumulative number of ionizing photons that have escaped into the IGM for a spherical region with smoothed overdensity $\delta_R$ and radius $R$, $\bar{n}_{\rm rec}$ is the mean cumulative number of recombinations per Hydrogen atom in the IGM, $ N_{\mathrm{H}}(\delta_R)$ is the number of neutral atoms in the region, and $f_{\rm coll}(>M_{\rm J})$ is the collapse fraction of halos above Jeans mass \citep{barkana2001}. Compared with the standard excursion-set model, here we add the ionizing photons consumed by minihalos inside the spherical region, $N_\xi$.

The cumulative number of ionizing photons
\begin{align}
N_{\rm ion}=&V
\int_{M_{\min}}^{M_{\rm max}}\,
\bigg[f_{\rm esc}\,N_{\gamma/\mathrm{b}}\,
f_*(M_\mathrm{h},z) \frac{\Omega_{\rm b}}{\Omega_{\rm m}}M_\mathrm{h}\, \nonumber \\
& \times f_{\rm duty}(M_\mathrm{h},z)
\frac{\mathrm{d}n}{\mathrm{d}M_\mathrm{h}}\bigl(M_\mathrm{h},z|\delta_{\mathrm{R}},M_{\mathrm{R}}\bigr)\bigg]\,\mathrm{d}M_\mathrm{h}
\label{eq:nion}
\end{align}
where $V$ is the region volume,  
$M_{\rm max}$ is the maximum mass of halos in the spherical region,
$f_{\mathrm{esc}}$ is the escape fraction, and $N_{\gamma/\mathrm{b}}$ is the number of ionizing photons per stellar baryon produced in their lifetime, for which we adopt $9{,}500$, provided  by \texttt{Starburst99} \citep{leitherer99,vzquez2005,leitherer2009,leitherer2014} for a Kroupa IMF \citep{kroupa2001} and metallicity  $Z=0.05Z_{\odot}$.
 The conditional mass function $\frac{\mathrm{d}n}{\mathrm{d}M_\mathrm{h}}\bigl(M_\mathrm{h},z|\delta_{\mathrm{R}},M_{\mathrm{R}}\bigr)$ is provided by the Extended Press-Schechter (EPS) formalism \citep{mo1996}.
Similar to {\tt 21cmFAST} \citep{mesinger2011}, we normalize the mean number density of ionizing photons  in the Universe to the values given by Sheth \& Tormen halo mass function.
Throughout this paper, we set $M_{\rm max}={\rm min}(10^{15}~M_\odot,0.9M_{\rm R})$, where $M_R$ is the mass of the spherical region.

The gas numbers in the spherical region with volume $V$ simply 
\begin{equation}
N_{\mathrm{H}}  = \frac{V}{\mu_{\mathrm{b}}m_\mathrm{H}}\bar{\rho}_{\mathrm{b}}\left(1+\delta_{R}\right),
\end{equation}
where
$\bar{\rho}_\mathrm{b}$ is the mean comoving baryon density of the Universe, $\mu_{\rm b}$ is the mean molecular weight and $m_{\rm H}$ is the mass of Hydrogen atom.

The recombination numbers $\bar{n}_\mathrm{rec}$ is solved from the  following differential equation {\citep{choudhury2025}}, 
\begin{equation}
\frac{\mathrm{d}n_{\mathrm{rec}}}{\mathrm{d}t}
= \chi_{\mathrm{He}}\, C_{\mathrm{HII}}\, n_\mathrm{H}\, X_{\mathrm{HII}}\,
\alpha_{\mathrm{A}}\bigl(T_{\mathrm{HII}}\bigr)\,(1+z)^{3},
\label{eq:nrec}
\end{equation}
where $\chi_{\mathrm{He}} \approx 1.08$ is the contribution of singly-ionized Helium to the free electron density, $n_\mathrm{H}$ is the comoving Hydrogen number density, $C_{\mathrm{HII}}$ is the IGM clumping factor \citep{shull2012}, \begin{equation} 
C_\mathrm{HII}=2.9[(1+z)/6]^{-1.1},
\end{equation}
and $\alpha_{A}$ is the Case A recombination coefficient. To simplify the calculation, we assume the IGM temperature  $T_{\mathrm{HII}}=  10^4\,\mathrm{K} $ that gives $\alpha_{\mathrm{A}} = 4.2 \times 10^{-13}$. 
The quantity of the $n_\mathrm{rec}$ at redshift $z$ is obtained simply by integrating Eq. (\ref{eq:nrec}) from $\infty$ to $z$:
\begin{equation}
n_{\mathrm{rec}}=\int_{\infty}^{z}\mathrm{d}z\;\frac{\mathrm{d}t}{\mathrm{d}z}\;
\frac{\mathrm{d}n_{\mathrm{rec}}}{\mathrm{d}t}.
\label{eq:nrecint}
\end{equation}

The number of photons consumed by minihalos
\begin{align}
N_{\mathrm{\xi}}(z| R,\delta_\mathrm{R}) = &    V
\int_{M_{\mathrm{J}}}^{M_{\mathrm{min}}} \bigg[ 
\xi(M_\mathrm{h},z)\, \frac{1}{\mu_{\rm b}m_{\rm H}}
\frac{\Omega_{\rm b}}{\Omega_{\rm m}}M_\mathrm{h} \,\nonumber \\ 
&  \times \frac{\mathrm{d}n}{\mathrm{d}M_\mathrm{h}} (M_\mathrm{h},z|\delta_{\mathrm{R}},M_{\mathrm{R}})\bigg]\,\mathrm{d}M_\mathrm{h},
\label{eq:minihalo_sink}
\end{align}
where $\xi(M_\mathrm{h},z)$ is the mean number of ionizing photons consumed per Hydrogen atom until a minihalo is completely photo-evaporated by the ionization front \citep{iliev2005}.

We use the {\tt 21cmFAST} \citep{mesinger2007,mesinger2011,park2019,Murray2020JOSS} to generate density fields for both fiducial  model and the C25 models, following the Zel'dovich approximation algorithm \citep{yb1970}. 
The density fields have box size 300 Mpc and 300$^3$ cells. We then apply the above excursion-set algorithm to the density fields, to generate the ionizing fields. This is also analogous to {\tt 21cmFAST}. Explicitly, We adopt a series of smoothing radii $R$ ranging from $50 \, \mathrm{Mpc}$ down to $1\, \mathrm{Mpc}$. For each cell, 
starting from the largest smoothing radius, we  check if the ionization condition Eq.~(\ref{eq:excursion minihalo}) is satisfied. If it is satisfied, then the cell is marked as ``ionized'', otherwise we move to smaller smoothing radius. If the cell remains neutral at the smallest radius, it is assigned a partial ionization fraction defined by the ratio of the LHS to the RHS of Eq.~(\ref{eq:excursion minihalo}). This accounts for contributions of ionized bubbles smaller than the minimum smoothing radius. Such a treatment is not crucial for the fiducial model, as the contribution of partially ionized cells  to the total ionization fraction is $\sim 0.15$. However, in the C25 model it is important, because the matter power spectrum enhancement at small scales largely boosts the number of minihalos and the ionization barrier. It trends to form more smaller ionized bubbles rather than fewer larger bubbles. For example, when $k_{\rm trans}=150\,\mathrm{Mpc}^{-1}$, the contribution to the total ionization fraction from partially ionized cells is $\sim 0.32$. 
We remind, however, that the distinction between partial ionization and full ionization depends on resolution. A partially ionized cell in a lower resolution box will be split into some fully ionized cells and some neutral cells if increasing the resolution. Nevertheless, the 21 cm power spectrum and the BSD at scales much larger than resolution will be less sensitive to this, but the integrated CMB scattering optical depth may be influenced.

We then calculate the 21 cm signal field from the ionization field. The 21 cm  brightness temperature \citep{fzh04}
\begin{align}
\delta T_{\rm b} (\delta_{\rm b},z)\approx & 27\, X_{\mathrm{HI}}(1+\delta_{\rm b})
\left(\frac{\Omega_{\rm b} h^2}{0.023}\right)
\left(\frac{0.15}{\Omega_{\rm m} h^2}\frac{1+z}{10}\right)^{1/2} \nonumber \\
& \times \left(\frac{T_{\mathrm{S}}-T_{\mathrm{CMB}}}{T_{\mathrm{S}}}\right)
\,\mathrm{mK},
\label{eq:21cm}
\end{align}
where $X_{\mathrm{HI}}$ is the neutral Hydrogen fraction, $T_{\mathrm{S}}$ is the Hydrogen spin temperature, $T_{\mathrm{CMB}}$ is the CMB temperature. Throughout this paper, we limit our investigation in the EoR when the IGM has been heavily heated and Ly$\alpha$ scattering has been saturated, it is reasonable to assume that $T_{\rm S} \gg T_{\rm CMB}$ \citep{HERA2023ApJ}.

\begin{figure*}
    \centering
    \includegraphics[width=\linewidth]{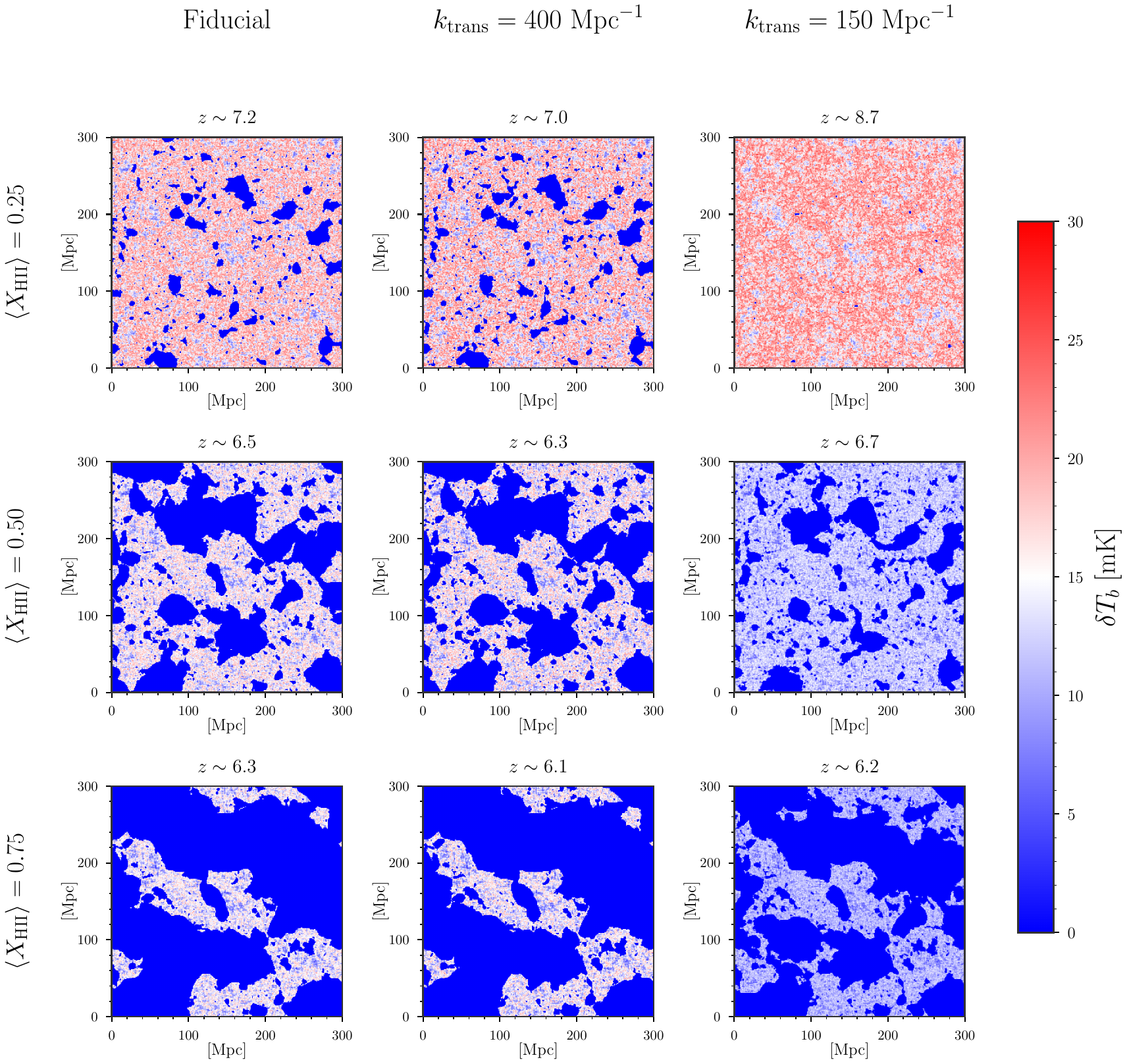}
    \caption{ 
The 21 cm fields for fiducial model and  C25 models with $k_\mathrm{trans}=400\,\mathrm{Mpc}^{-1}$ and $k_\mathrm{trans}=150\,\mathrm{Mpc}^{-1}$. The first row displays the results for mean ionization fraction $\langle X_{\rm HII}\rangle=0.25$, the second row $\langle X_{\rm HII}\rangle=0.50$ and the third row $\langle X_{\rm HII} \rangle=0.75$. The redshift of each panel is given above the top $x$-axis.
 }
    \label{fig:xxifield}
\end{figure*}

In Figure \ref{fig:xxifield} we show the 21 cm signal fields at $\langle X_\mathrm{HII} \rangle =$ ($0.25$, $0.50$, $0.75$) for the fiducial model and the C25 models with $k_{\rm trans}=400 \,\mathrm{Mpc}^{-1}$ and $k_{\rm trans}=150 \,\mathrm{Mpc}^{-1}$ respectively, 
we set $f_{\rm esc}=0.15$.
In the C25 model, these two $k_{\rm trans}$ values represent two different typical situations. For $k_{\rm trans}=400$ Mpc$^{-1}$, only minihalos are boosted. So it purely enhances the consumption of ionizing photons, but the production of ionizing photons and the galaxy UV LFs are same to the fiducial model.

For $k_{\rm trans}=150$ Mpc$^{-1}$, however, both minihalos and galaxies are boosted. This is a more complicated situation as the net effects depend on the competition between the enhancements on consumption and on production of ionizing photons. 
Moreover, according to Figure \ref{fig:uvlf_grid}, although the observed UV LFs allow the $k_{\rm trans}$ be smaller than 150 Mpc$^{-1}$, we will see in the next section that, if $k_{\rm trans} \lesssim $ 150 Mpc$^{-1}$, for $f_{\rm esc}=0.15$ it may produce too many ionizing photons and  break the constraints set by CMB scattering optical depth.

From Figure \ref{fig:xxifield}, we see that

for the C25 model with $k_\mathrm{trans}=400\,\mathrm{Mpc}^{-1}$ the 21 cm signal field is almost same to fiducial model as long as $\langle X_{\rm HII} \rangle$ is the same. However, for $k_\mathrm{trans}=150\,\mathrm{Mpc}^{-1}$, the size of ionized regions are much smaller than the `fiducial model, even the mean ionization fraction $\langle X_{\rm HII} \rangle$ is the same.

Particularly, in this case when $\mean{x_{\rm HII}}=0.25$ (see the top right panel of Figure \ref{fig:xxifield}), most of the contributions are from cells marked as ``partially ionized''. However, we remind that partial ionization does not necessarily mean the mixed neutral and ionized atoms, instead, it actually refers to the volume fraction of fully ionized bubbles in that cell. The real partial ionization front of UV ionizing photons is quite thin and only occupies a small fraction in the volume. \citep[e.g.,][]{iliev2006MNRAS.371.1057I}

\section{Results}
\label{sec:result}

\subsection{The reionization history}

The Thompson scattering optical depth to the CMB 
\begin{equation}
    \tau_e(z)=c\sigma_{\rm T}\bar{n}_\mathrm{H} \int_0^z 
    \frac{(1+z^\prime)^2\mathrm{d}z^\prime}{H(z^{\prime})}\langle X_\mathrm{HII}\rangle \,
    \left( 1+\eta\frac{Y}{4X}\right),
\end{equation}
where, $c $ is the speed of light, $\sigma_{\rm T}$ is the Thomson cross-section, $\bar{n}_{\rm H}$ is the mean comoving number density of Hydrogen,   $H(z)$ is the Hubble parameter, and $X$ and $Y$ are the primordial mass fractions of Hydrogen and Helium, respectively. The $\eta=1$ ($z>3$) or $\eta=2$ ($z<3$) measures the degree of Helium ionization based on observations \citep{shull2010,shull2012,shapiro2004}. 

\begin{figure}[ht!]
\centering
\includegraphics[width=0.4\textwidth]{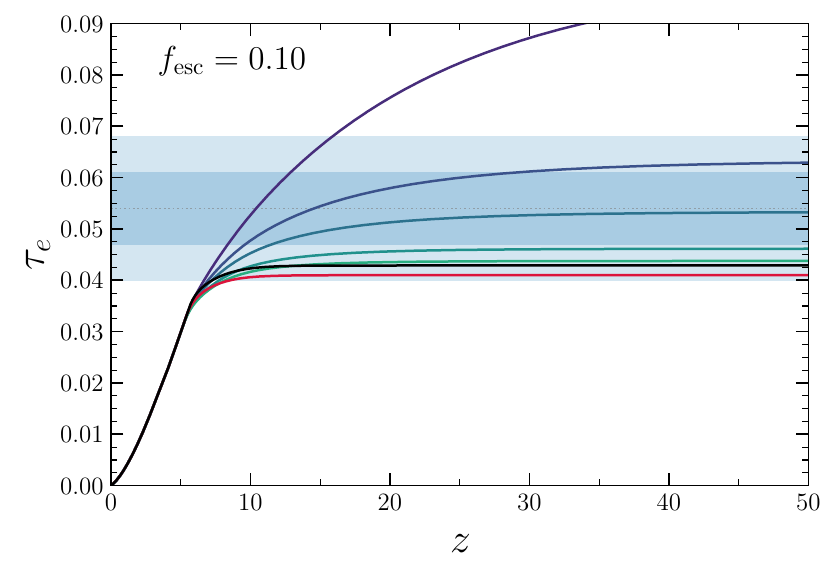}\\

\includegraphics[width=0.4\textwidth]{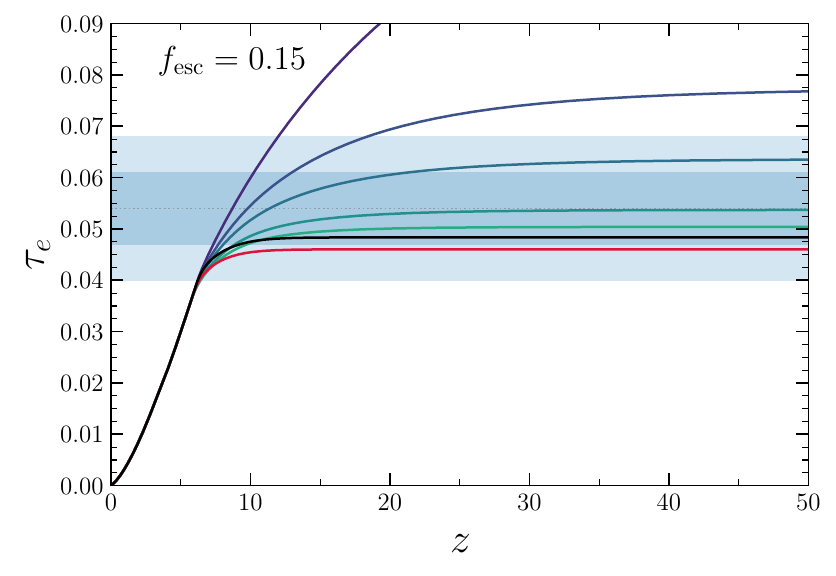}\\

\includegraphics[width=0.4\textwidth]{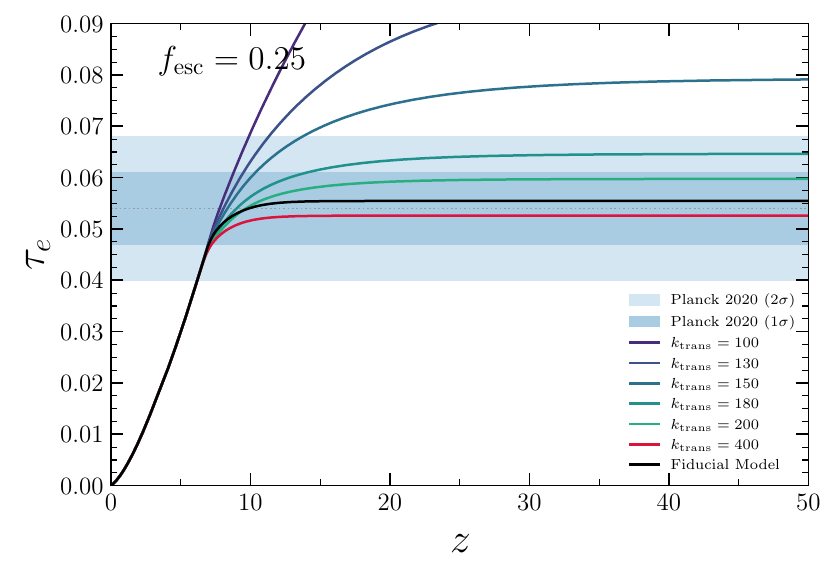}
\caption{CMB scattering optical depth as a function of $z$ for the fiducial model and C25 models with different $k_{\rm trans}$ values. The shaded filled regions are observational data and 1$\sigma$ and 2$\sigma$ uncertainties of Planck18 \citep{planck18}.
}
\label{fig:depth}
\end{figure}

In Figure \ref{fig:depth} we plot the $\tau_{\rm e}$ in fiducial model and the C25 models with different $k_{\rm trans}$ values, for different $f_{\rm esc}$, compared with the observed $\tau_e = 0.054 \pm 0.007$ \citep{planck18}.  In all cases the fiducial model always has optical depth consistent with observations. Moreover, when $f_{\rm esc}=0.10$, models with $k_{\rm trans} \lesssim 130$ Mpc$^{-1}$ are excluded by optical depth observations at 2$\sigma$ level. While when $f_{\rm esc}=0.15$, models with $k_{\rm trans} \lesssim 150$ Mpc$^{-1}$, and when $f_{\rm esc}=0.25$, models with $k_{\rm trans} \lesssim 180$ Mpc$^{-1}$, are excluded.

\begin{figure}[ht!]
\centering
\includegraphics[width=0.4\textwidth]{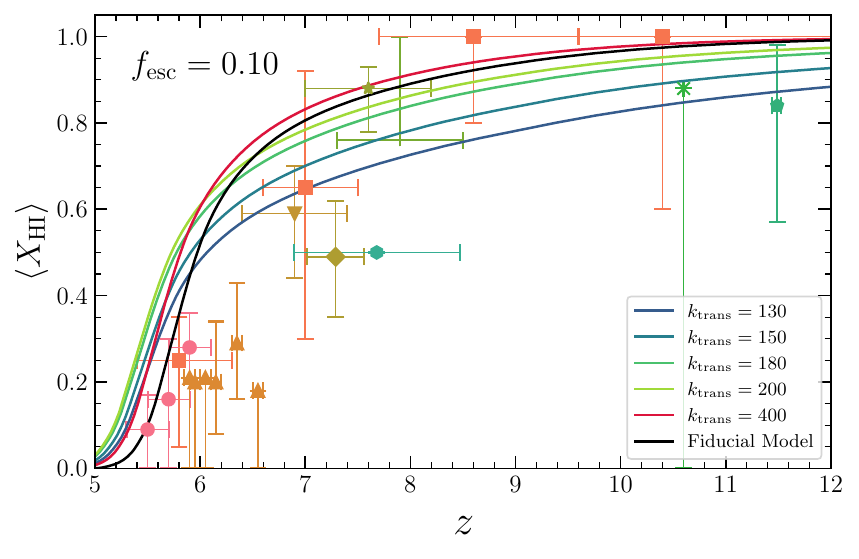} \\

\includegraphics[width=0.4\textwidth]{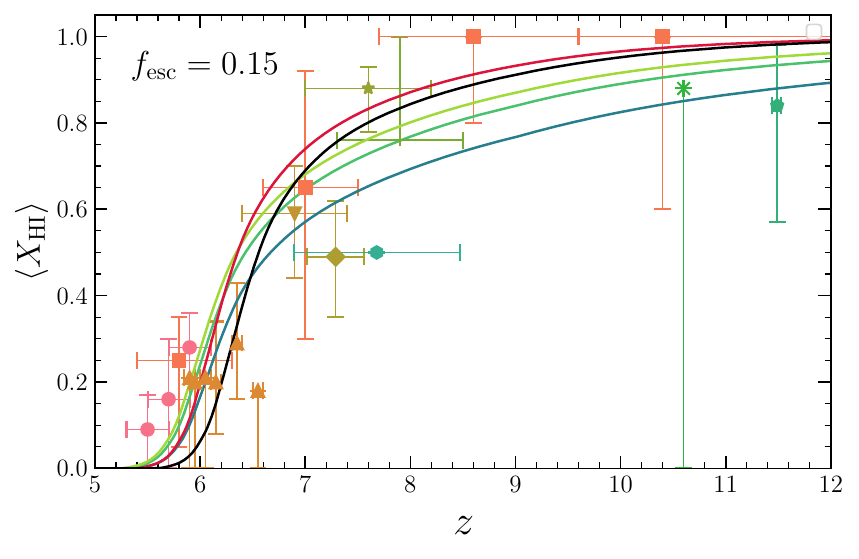} \\

\includegraphics[width=0.4\textwidth]{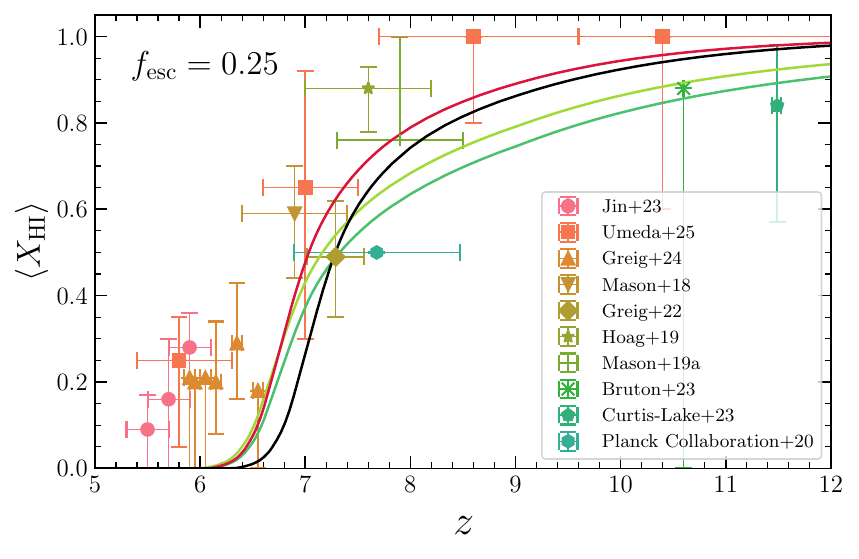}
\caption{Reionization history in fiducial model and C25 models with different $k_{\rm trans}$ values, and in observations, including Lyman-$\alpha$ emission equivalent widths in galaxies 
\citep{mason2018, mason2019, hoag2019, bruton2023}, damping wing absorption in quasar and galaxy spectra \citep{curtis2023, greig2022, greig2024, umeda2025}, and dark pixel fractions in the Lyman-$\alpha$/Lyman-$\beta$ forests \citep{jin2023}.
}
\label{fig:eor}
\end{figure}

In Figure \ref{fig:eor} we plot the reionization history in the fiducial model and the C25 models with different $k_{\rm trans}$ values, together with the neutral fractions in various observations \citep{mason2018,mason2019, hoag2019,bruton2023,curtis2023, greig2022,greig2024, umeda2026,jin2023}. 
For $f_{\rm esc}=0.10$ and 0.15, all models with $k_{\rm trans} \gtrsim 130$ Mpc$^{-1}$ are consistent with observations. For $f_{\rm esc}=0.25$, it seems that reionization ended a bit earlier than, but still marginally consistent with, observations.

It is interesting to note that reionization history provides more information than the integrated variable $\tau_e$. For C25 models with $130 \lesssim k_{\rm trans} \lesssim 200$ Mpc$^{-1}$, the boost on ionizing photons production dominates over the boost on consumption, so the reionization is promoted at the early stage of the EoR ($\langle X_{\rm HII}\rangle\lesssim 0.3 - 0.6$), compared with the fiducial model. However, at the late stage of the EoR ( $\langle X_{\rm HII}\rangle \gtrsim 0.3 - 0.7$), when most of regions have been ionized, boost on ionizing photons consumption in minihalos dominates over the production in halos $> M_{\rm min}$, as a results the reionization is delayed.

In addition to the UV LFs, the CMB scattering optical depth and the reionization history put stricter constraints on the $k_{\rm trans}$ values in the C25 models. However, it is still infeasible to fully distinguish them, as the optical depth is an integration, while the observations of ionized fractions are still highly uncertain. We will investigate the possible constraints  from 21 cm observations in following subsections.

\subsection{The 21 cm power spectrum and detectability}

\begin{figure*}
\centering
\includegraphics[width=0.47\textwidth]{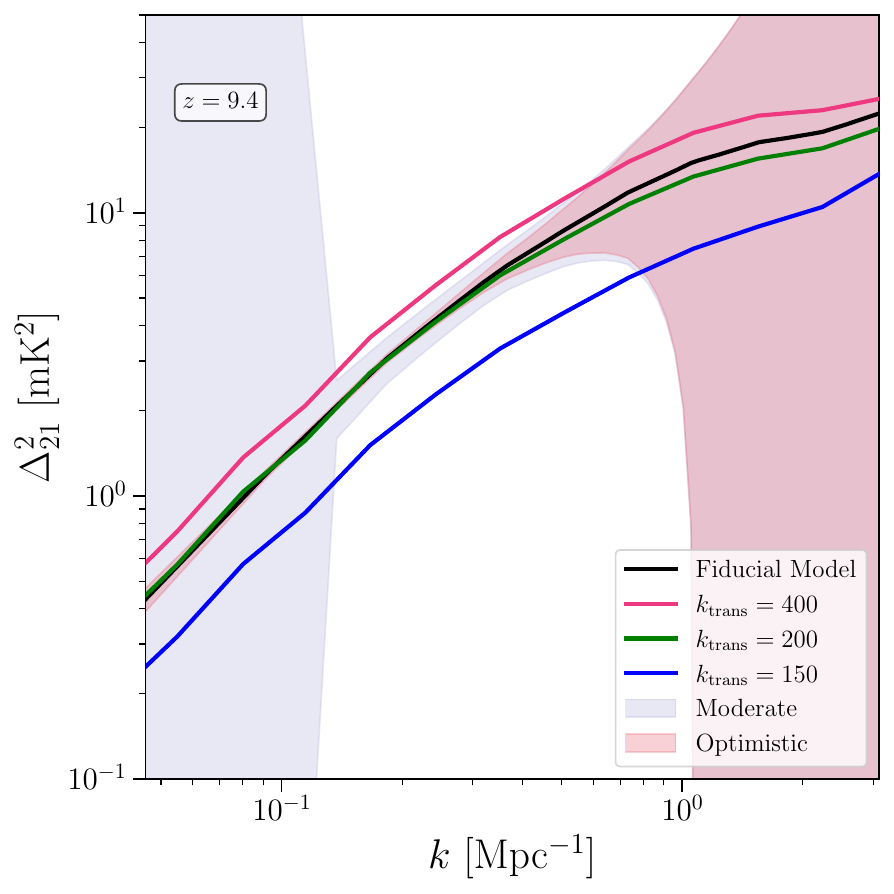}
\hfill
\includegraphics[width=0.47\textwidth]{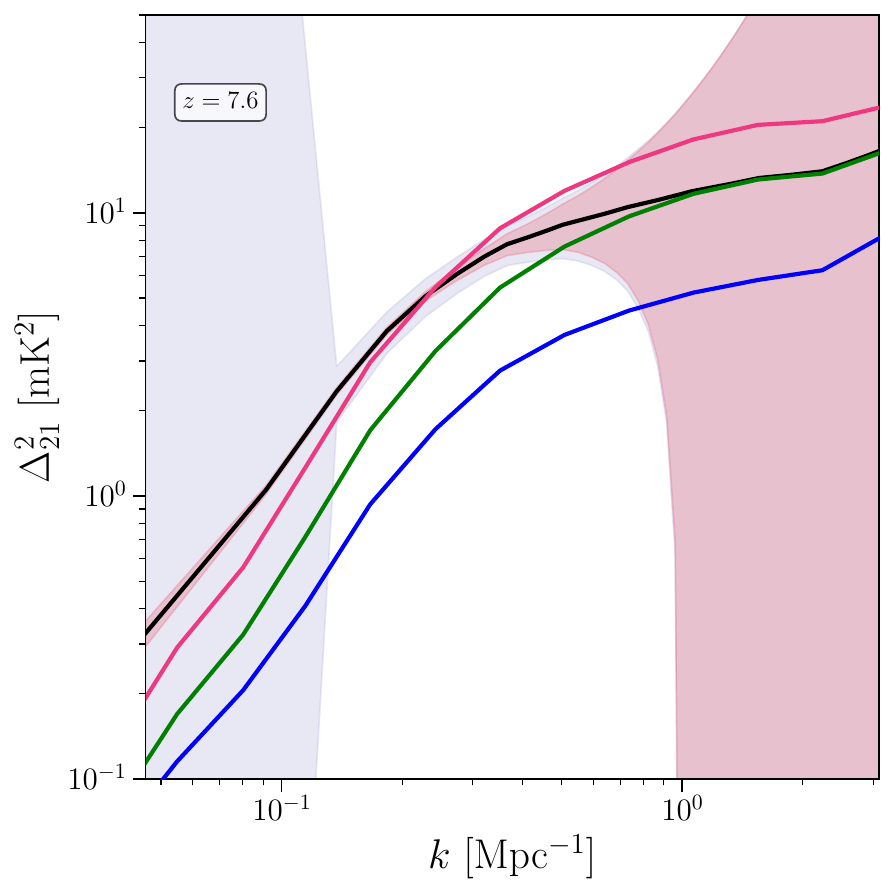}

\includegraphics[width=0.47\textwidth]{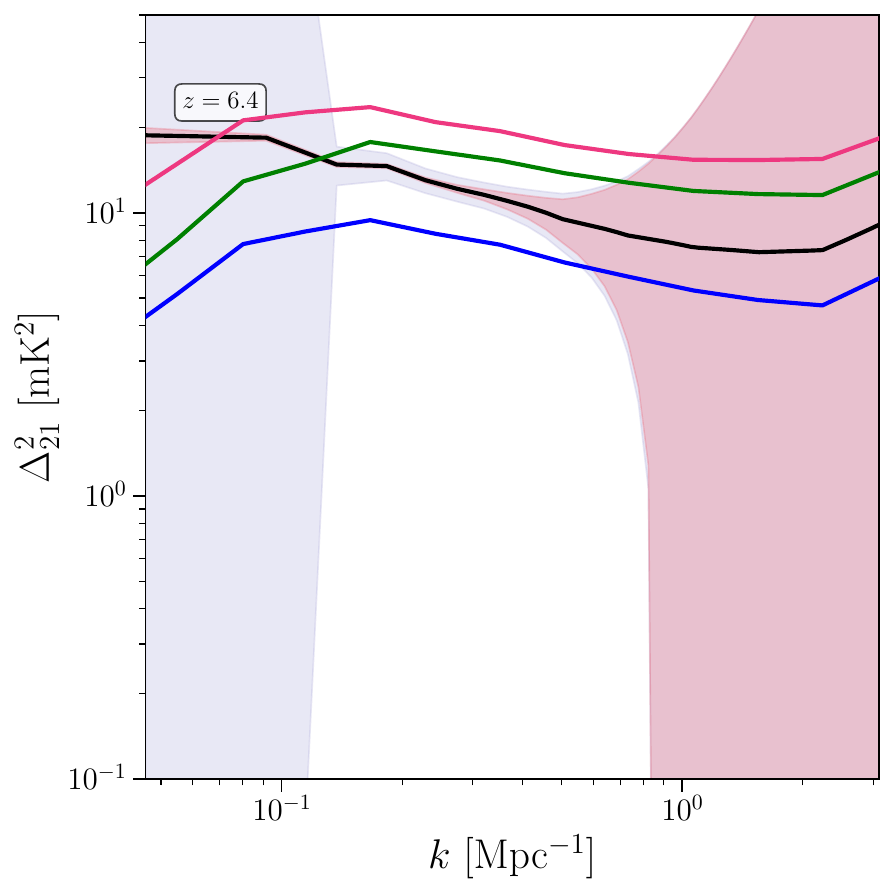}
\hfill
\includegraphics[width=0.47\textwidth]{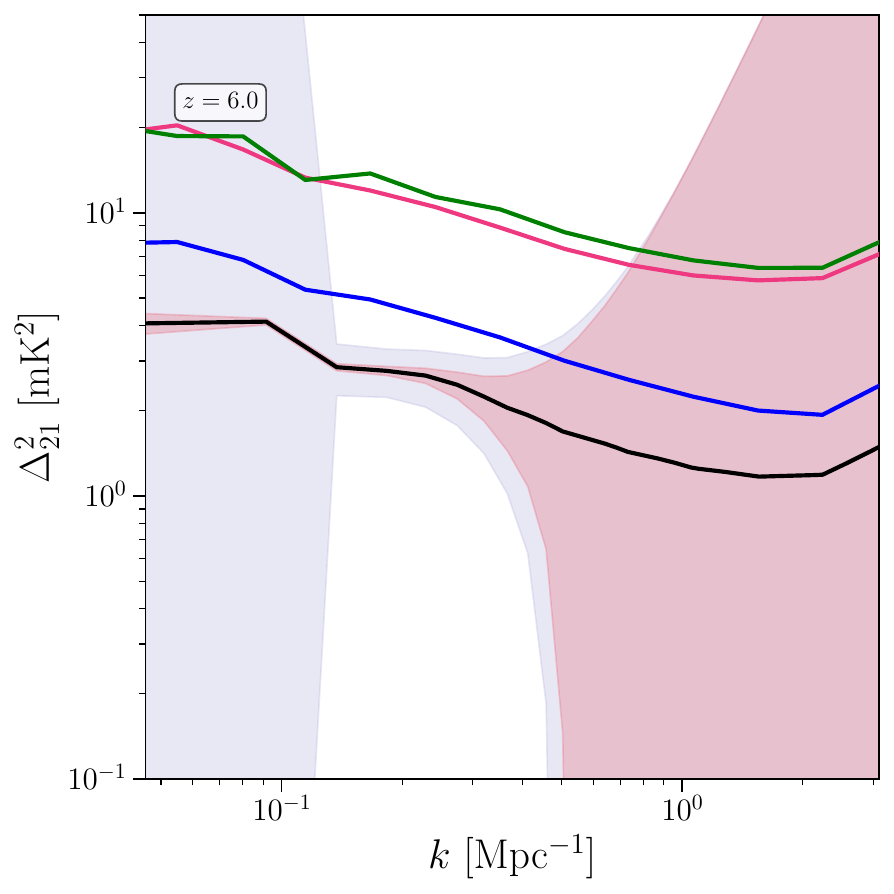}

\caption{
The 21 cm power spectra in fiducial model and C25 models at different redshifts and the uncertainties for SKA-low AA* observations.
}
\label{fig:21cm}
\end{figure*}

Since in Figure \ref{fig:eor}, for $f_{\rm esc}=0.25$ the reionization history is a bit earlier than observations, we adopt $f_{\rm esc}=0.15$ in following calculations. We will no longer consider models with $k_{\rm trans} \lesssim 150$ Mpc$^{-1}$, as they are ruled out by the observed $\tau_e$ when $f_{\rm esc}=0.15$. In Figure \ref{fig:21cm} we show the 21 cm power spectra in the fiducial  model and the C25 models with different $k_{\rm trans}$ values, at redshifts $z=9.4, 7.6, 6.4$ and 6.0 respectively. We also plot the expected observational uncertainty levels for the SKA-low AA*. The observational uncertainties are calculated by using the {\tt 21cmSense} \citep{pober2013,pober2014,murray2024}. We employ the SKA-low AA* array within a 700 m radius, 8 MHz bandwidth and total 1080 hour integration. 
The \texttt{21cmSense} provides three types of foreground removal modes: ``Pessimistic'', ``Moderate'',  and ``Optimistic''. Different foreground removal modes in \texttt{21cmSense} provide estimates of the observational limits for  the foreground wedge, which determines the boundary of $k$-modes in the $k_\parallel$--$k_\perp$ Fourier space. 
We show the results for moderate and optimistic removal modes in Figure \ref{fig:21cm}.

At $z=9.4$, the 21 cm  power spectrum for $k_{\rm trans}=150$ Mpc$^{-1}$ is smaller than that of the fiducial  model, while the power spectrum for $k_{\rm trans}=400$ is larger. This is because in the $k_{\rm trans}=150$ Mpc$^{-1}$ model the ionizing sources are a bit more than the fiducial $\Lambda$CDM model, although the minihalos are also more abundant, the ionization is a bit promoted and the ionization fraction is a bit higher. In contrast, for the $k_{\rm trans}=400$ Mpc$^{-1}$ model, the ionizing sources are same to fiducial  model, but minihalos consume more ionizing photons. The reionization is delayed and the ionization fraction  is a bit lower,  see Figure \ref{fig:eor}. The above difference in 21 cm power spectrum is larger than uncertainties of the SKA-low AA* observations, in principle the models are distinguishable. For the $k_\mathrm{trans}=200$ Mpc$^{-1}$ model, however, the signal is very close to the fiducial model. We check that this is because in this model and at this moment, the enhancement of ionizing photons production in halos with $T_{\rm vir} \gtrsim 10^4$ K and consumption in minihalos coincidentally counteract each other.

At $z=7.6$, the difference between {the fiducial model} and {the C25 model} with $k_{\rm trans}=400$ Mpc$^{-1}$ becomes smaller, 
the $k_{\rm trans}=400$ Mpc$^{-1}$ model has a smaller 21 cm power spectrum at large scale ($k\lesssim 0.3$ Mpc$^{-1}$), but larger at small scale. However, this difference is comparable to SKA-low AA* observational uncertainties for moderate foreground removal model, so hard to distinguish. For optimistic foreground removal model the uncertainties are smaller at larger scales $k\lesssim 0.1$ Mpc$^{-1}$, the difference are distinguishable. The $k_{\rm trans}=150$ Mpc$^{-1}$ model is still a smaller than that of fiducial model, and is distinguishable. For $k_\mathrm{trans}=200$ Mpc$^{-1}$ model, the 21 cm power spectrum is lower than the fiducial model at large scale, and can be distinguished in both the moderate and optimistic foreground removal modes.

At $z=6.4$, the difference between the $k_\mathrm{trans}=400$ Mpc$^{-1}$ model and the fiducial model becomes larger, the $k_\mathrm{trans}=400$ Mpc$^{-1}$ model is larger than fiducial model at $k\gtrsim 0.06\, \mathrm{Mpc}^{-1}$. However the difference $k_\mathrm{trans}=150$ Mpc$^{-1}$ model and the fiducial model becomes smaller. This is because in these two models the reionization process is delayed by minihalos, see Figure \ref{fig:eor}. The difference between the three models is fully distinguishable for both the moderate and optimistic foreground removal models. For the $k_\mathrm{trans}=200$ Mpc$^{-1}$ model, at this redshift the 21 cm power spectrum is larger than the fiducial model at $k\gtrsim 0.1$ Mpc$^{-1}$, however still smaller than the fiducial model at larger scale.

At $z=6.0$, the late stage of the EoR,  the 21 cm power spectra in all C25 models are larger than the fiducial model, because their reionization is delayed, and the differences are detectable for the SKA-low AA*. 

We caution that here we compare the 21 cm power spectra in different models at the same redshift. Different models have different reionization history, although the integrated $\tau_e$ are similar. Therefore the difference in 21 cm power spectra arises  from both different $\mean{X_{\rm HII}}$ and different ionized bubble morphology.

\subsection{The bubble size distribution and detectability}

In this subsection we investigate the BSD. BSD is an intuitive indicator for morphology of ionization field \citep{giri2018}. Ionized bubbles must be identified in tomographic images, while tomographic imaging observation is far beyond the scope of SKA-low AA*. Here we base our investigations on an assumed telescope with sensitivity much larger than the SKA-low AA*.

\begin{figure*}[ht!]
\includegraphics[width=0.3\textwidth]{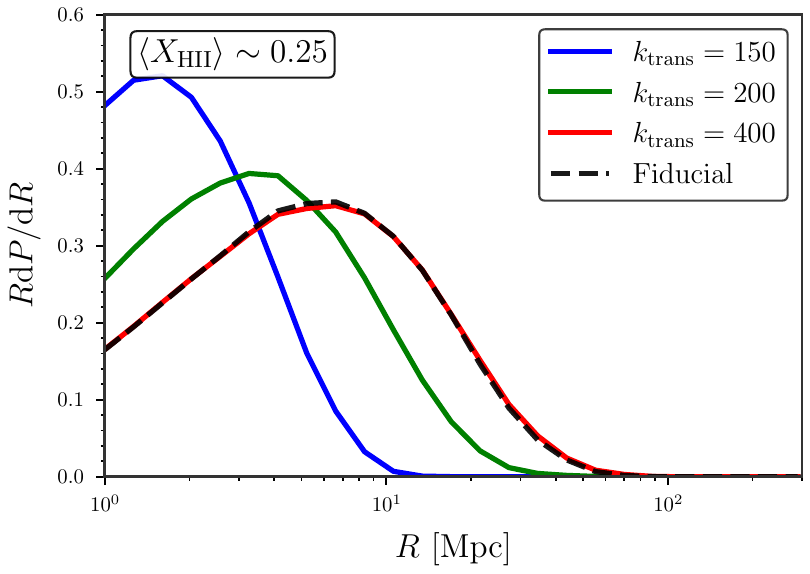}
\includegraphics[width=0.3\textwidth]{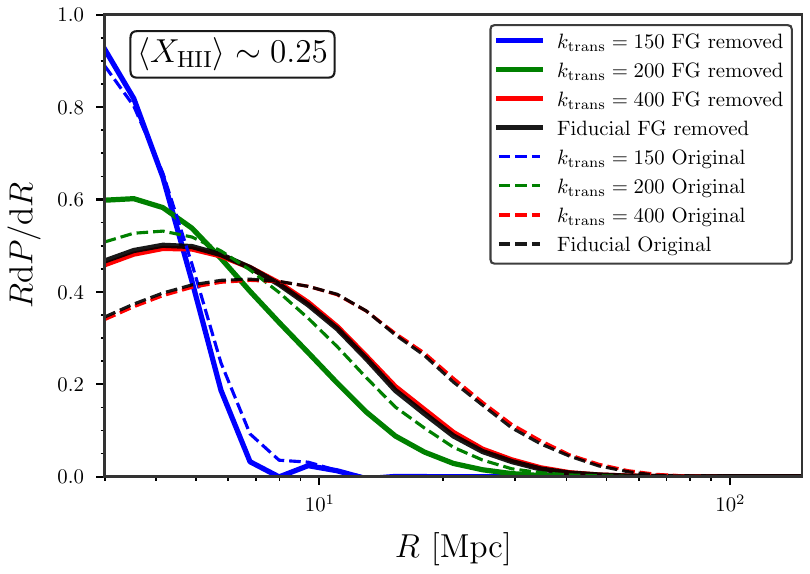}
\includegraphics[width=0.3\textwidth]{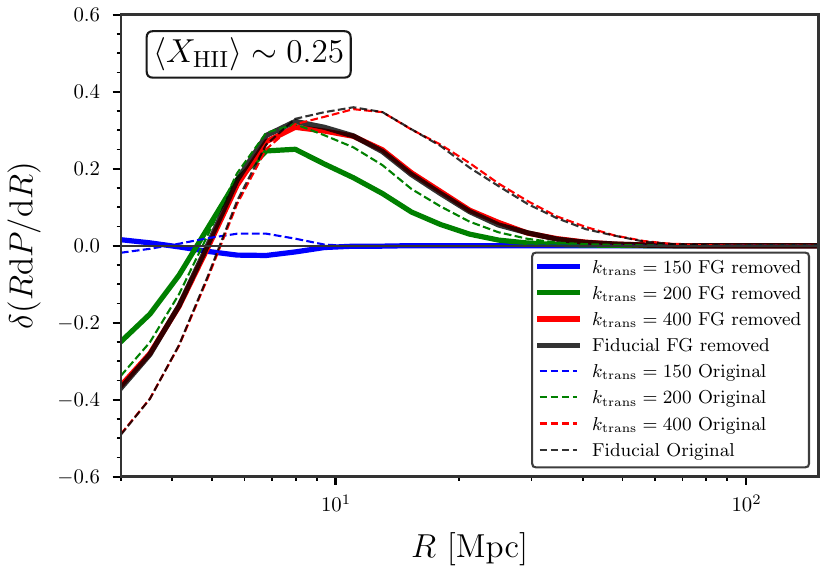} \\
\hfill
\includegraphics[width=0.3\textwidth]{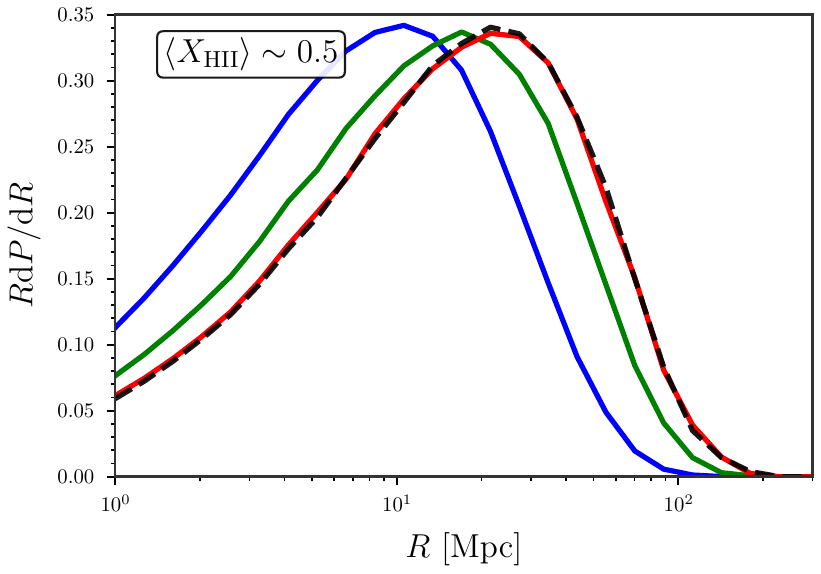}
\includegraphics[width=0.3\textwidth]{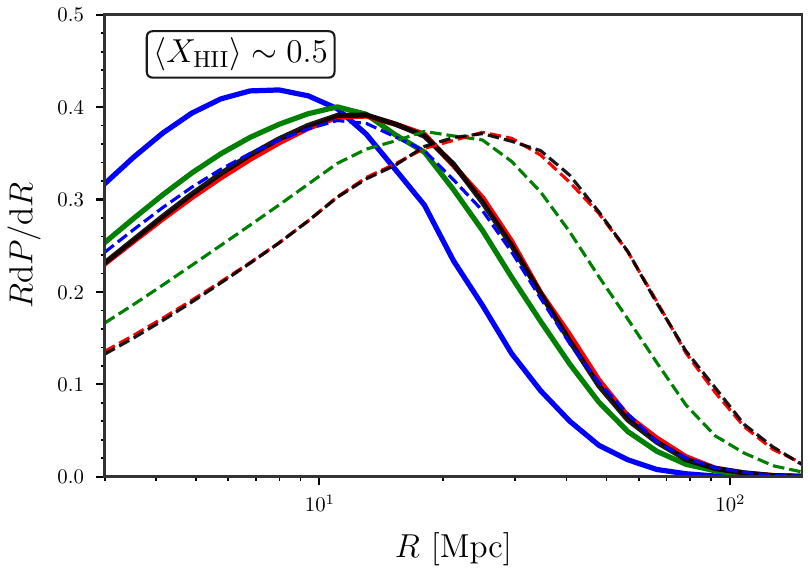}
\includegraphics[width=0.3\textwidth]{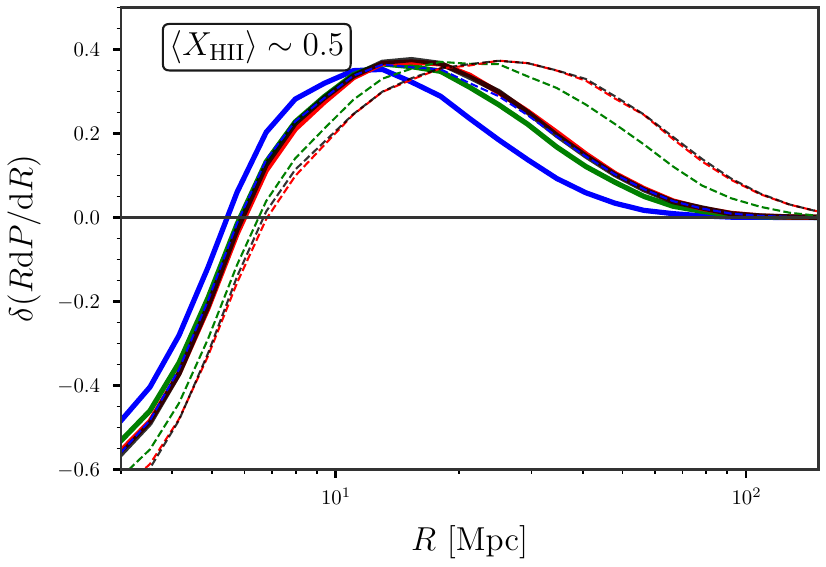}\\
\hfill
\includegraphics[width=0.3\textwidth]{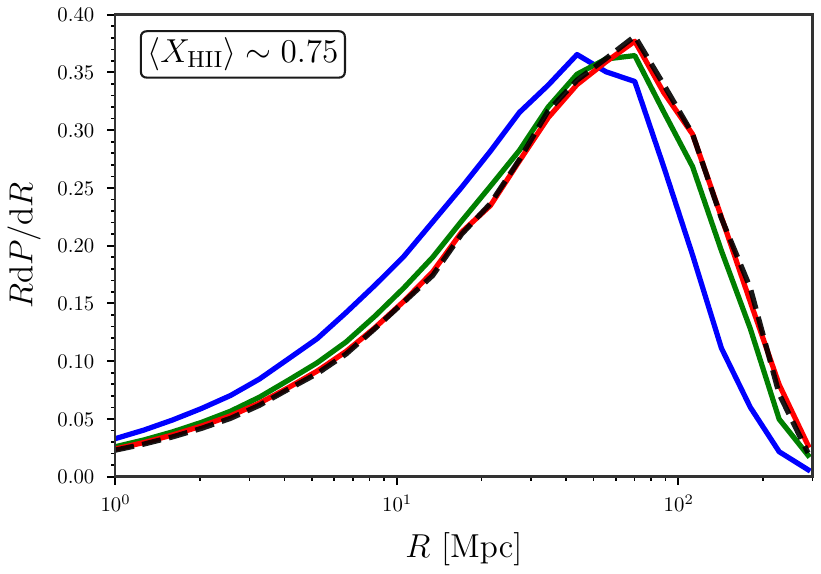}
\includegraphics[width=0.3\textwidth]{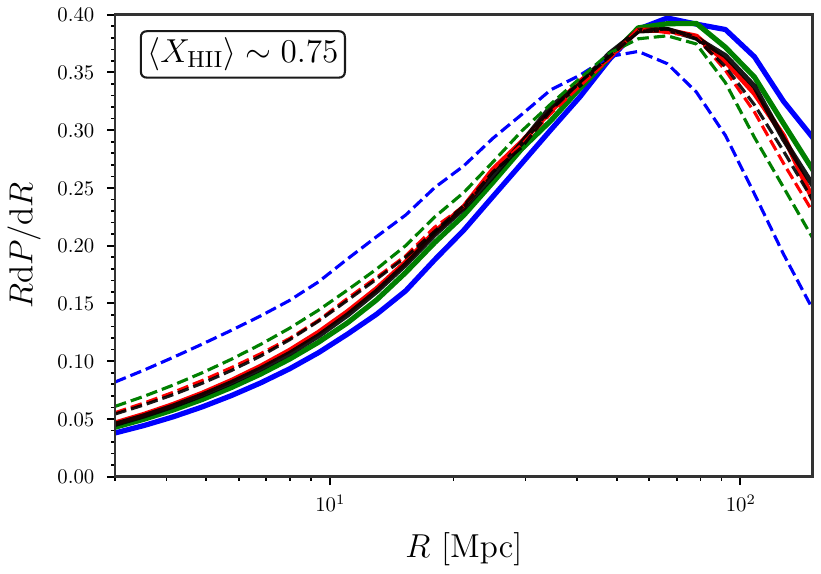}
\includegraphics[width=0.3\textwidth]{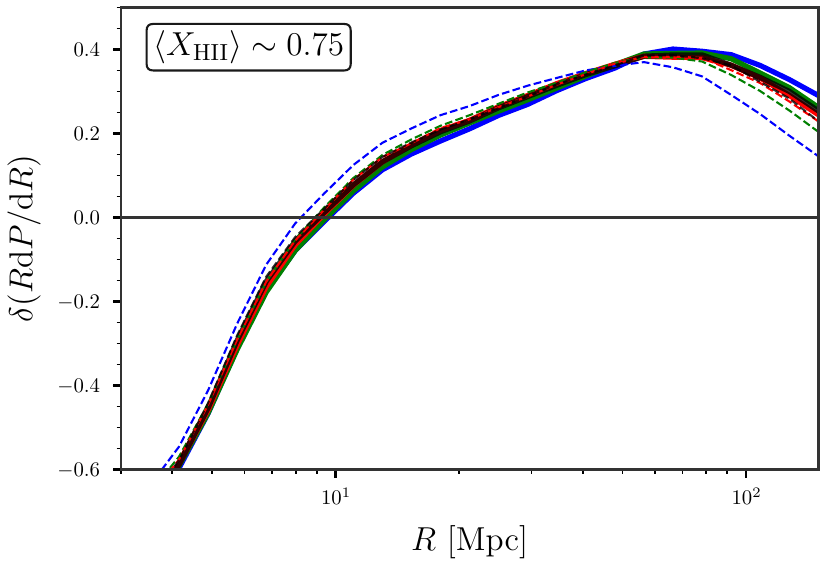}
\hfill
\caption{
The BSD for the fiducial model and C25 models. {\it Top row:} $\langle X_{\rm HII}\rangle\sim 0.25$. {\it Middle row:} $\langle X_{\rm HII}\rangle\sim 0.50$. {\it Bottom row:} $\langle X_{\rm HII}\rangle\sim 0.75$.
{\it Left column:} the BSD is directly measured from the ionization fields.
{\it Middle column:} the BSD$_{21}$ is measured from the 21 cm fields, in the absence/presence of foreground contamination.
{\it Right column:} we subtract the  ``BSD'' of pure neutral fields from the middle column. 
}
\label{fig:bsd}
\end{figure*}

There are different methods to identify bubbles, such as distance transform (DT, \citealt{zahn2007}), mean free path (MFP, \citealt{mesinger2007}), and friends-of-friends (FoF, \citealt{iliev2006,friedrich2011}). Here we employ the MFP method provided in the code {\tt Tools21cm} \citep{giri2018,giri2020} to  derive the  BSD. This method first defines the ionized region, then randomly selects an ionized point and a direction, and then records the distance to the neutral boundary along that direction. We adopt $X_{\rm HII}=0.50$ as the threshold ionization fraction for ionized bubbles \citep{mesinger2007} and construct $10^7$ direction lines.
The results for the fiducial model and C25 models are shown in left column of Figure~\ref{fig:bsd}, where the BSDs at stages with same  $\mean{X_{\rm HII}}$ are plotted in the same panel.

We find that, in different models BSDs are rather different even for the same $\mean{X_{\rm HII}}$. For C25 model with $k_{\rm trans}=400$ Mpc$^{-1}$, the BSD is almost identical to the fiducial model at all $\mean{X_{\rm HII}}$s. It implies  that although in this model the minihalos are more abundant than fiducial model, they just globally shift the reionization history, rather than change the morphology of ionization field. For $k_{\rm trans}\lesssim 400$ Mpc$^{-1}$, the BSD deviates from the fiducial model. At $\mean{X_{\rm HII}}\sim 0.25$ the peak of BSD in fiducial model is $R\sim 8$ Mpc, while in $k_{\rm trans}=200$ Mpc$^{-1}$ and 150 Mpc$^{-1}$ models, the peaks shift to $R\sim 4$ Mpc and $R\sim 1.3$ Mpc respectively. Indeed, in C25 models the ionized bubbles tend to fragment into smaller ones since minihalos intercept the propagation of ionizing photons. However, with the increasing of $\mean{X_{\rm HII}}$, the BSDs in C25 models and in fiducial models trend to approach each others. At $\mean{X_{\rm HII}}\sim 0.75$, almost all BSDs have peak $R\sim 50$ Mpc.
From this, we conclude that the best stages for detecting the  small-scale enhancement of matter power spectrum is the early and mid EoR.
 
Ionization fields are not directly observable in practice, we also identify bubbles from mock 21 cm sky maps. To image the 21 cm fields rather than just measure the power spectrum, it requires a telescope with sensitivity much higher than SKA-low AA*. We assume observations have angular resolution $\sim 67^{\prime\prime}$, frequency resolution $\sim 0.175$ MHz, and noise level $\sim 2.8$ mK. We add foreground from the {\tt GSM} sky model \citep{gsm08,gsm16} and Gaussian random noise to the 21 cm fields, to generate the mock samples. We then remove the foreground using the Singular Value Decomposition (SVD) algorithm \citep{yue2015}.  We adopt temperature 0.5$\langle X_{\rm HI} \rangle$ times of the mean 21 cm brightness temperature as the threshold for bubbles, and then apply the MFP method to the foreground subtracted 21 cm sky maps. The size distributions of bubbles identified from 21 cm fields, BSD$_{21}$, are shown in the middle column of Figure \ref{fig:bsd}. We find that for our mock samples, the influence of the foreground contamination is modest. The BSD$_{21}$ in different models are clearly distinguishable at the early and mid EoR. However, if the noise level is much larger, then it may dominate over the bubble signal. 

Since $\delta T_b \propto (1+\delta_b) X_{\rm HI}$, so the BSD$_{\rm 21}$ derived from 21 cm fields contains not only the real ionized bubbles, but also coadjacent low-density regions miss-identified as ``bubbles''.  This may attenuate the  bubble information in ${\rm BSD}_{21}$, as for both fiducial model and the C25 models, the large-scale density fields are almost the same. For this reason, we generate reference 21 cm fields by setting $X_{\rm HI}=1$ everywhere, then derive $\delta {\rm BSD}={\rm BSD}_{21}-{\rm BSD}_{\rm 21,ref}$.  The results are shown in the left column of Figure \ref{fig:bsd}. We believe that such a quantity could be a more useful indicator for telling the morphology information of ionization fields.

\section{summary and discussion}

\label{sec:summary}

\subsection{summary}

We investigated the impacts of small-scale enhanced power spectrum on reionization morphology and 21 cm signal. In such models, the number of faint galaxies and minihalos would be much more abundant than in the regular $\Lambda$CDM model, so the effects on reionization depends on the competition between the ionizing photons production by faint galaxies, and consumption by minihalos. However, the observed UV LFs, CMB scattering optical depth, and the reionization history exclude models with large boost on the ionizing photons production, i.e. models with $k_{\rm trans} \lesssim 100-130$ Mpc$^{-1}$. 

For models with $ 100-130~{\rm Mpc^{-1}} \lesssim k_{\rm trans} \lesssim 400$ Mpc$^{-1}$, both the ionizing photons production in faint galaxies and consumption in minihalos are boosted, reionization process is promoted at the early EoR, but delayed at the late EoR. These effects bring us features on the 21 cm power spectrum. Compared with the fiducial model, the power spectrum is smaller at early EoR and larger at the late EoR. The differences are detectable for upcoming SKA-low AA*. These effects also result different BSDs at the same $\mean{X_{\rm HII}}$.  The typical bubble size is obviously smaller than the fiducial model, and the difference is detectable from further telescopes that are able to do imaging observations.

For models with $k_{\rm trans} \gtrsim 400$ Mpc$^{-1}$, boost on ionizing photons production is negligible, therefore the impacts on reionization mainly rely on minihalos that purely consume ionizing photons, as a result the reionization process is delayed and the ionized bubbles fragment into smaller ones. The difference of the 21 cm power spectrum between such a model and fiducial model is distinguishable for the SKA-low AA*. However, the BSD is almost the same as long as the $\mean{X_{\rm HII}}$ is the same. Therefore, combining the results of power spectrum and BSD would be useful for constraining the $k_{\rm trans}$ models.

\subsection{discussion}

We assume that all the minihalos must be photo-evaporated in the ionized bubbles, otherwise reionization is not able to process. This would overestimate the role played by minihalos during the EoR. Suppose the radius of the minihalo with mass $M_{\rm mh}$ is $R_{\rm mh}$ \citep{furlanetto2005}, then the cross-section is $\pi R^2_{\rm mh}$. If the IGM is transparent to the ionizing photons and only minihalos resist them, the mean free path of an ionizing photon is
\begin{equation}
l_\mathrm{mh} = \frac{1}{\int_{M_\mathrm{J}}^{M_{\rm min}} \pi R_\mathrm{mh}^2 \frac{\mathrm{d}n}{\mathrm{d}m}(M_\mathrm{mh},z) \,\mathrm{d}M_\mathrm{mh}}.
\end{equation}
At $z=6$, for C25 model with $k_\mathrm{trans}=150 \,~\mathrm{Mpc}^{-1}$, the $l_\mathrm{mh}=1.86\,\mathrm{Mpc}$;  for $k_\mathrm{trans}=400 \,~\mathrm{Mpc}^{-1}$, the $l_\mathrm{mh }=1.12\,\mathrm{Mpc}$; for the fiducial model $l_\mathrm{mh}=2.14\,~\mathrm{Mpc}$. The fraction of ionizing photons consumed by minihalos to a distance $R$ from a source is $f_\mathrm{a}= 1 - \exp(- R/l_\mathrm{mh})$. That is to say, in a spherical bubble with radius equal to $3\,~\mathrm{Mpc}$, the fraction of  ionizing photons consumed by minihalos for these three models are  0.80, 0.93, and 0.75 respectively. According to this estimation, we believe that our assumptions should be reasonable at least  for bubbles with $R \gtrsim 3$ Mpc.

In our paper, we use the Jeans mass as the lower limit of minihalos that can preserve gas content. However, it has been pointed out that, filtering mass is more accurate as it accounts for the evolution history of the IGM temperature \citep{Gnedin1998MNRAS.296...44G}. Actually, the gas content in smaller minihalos is a complicated problem. It has been investigated by many analytical models (e.g. \citealt{Oh2003MNRAS.346..456O}) and numerical simulations (e.g. \citealt{chan2024}).  They are basically consistent with each other. In \citet{chan2024}, the filtering mass is approximately $2\text{--}8$ times the Jeans mass. However, we check that if we use this filtering mass, in the C25 model with $k_{\rm trans}=150$ Mpc$^{-1}$, the consumption of ionizing photons just changes negligibly. This is because in this model,  $M_{\rm h}^{\rm bump}$ is much larger than the Jeans mass or filtering mass, therefore most of ionizing photons are consumed in minihalos $\gg$ the lower limit.

We always assume the matter power spectrum evolve linearly at all scales. However, at small scales the non-linear growth would be important even at redshifts as high as the EoR \citep{Iliev2002ApJ...572L.123I,smith2003MNRAS.341.1311S,takahashi2012ApJ...761..152T}. Generally, non-linear effect would boost the matter power spectrum and enhance the formation of minihalos. It is quite likely that if we involve this effect in our model, it will strengthen our conclusions. Moreover, there are other effects such as the dark matter - baryon relative streaming motion (e.g. \citealt{naoz2013ApJ...763...27N}) and the dark matter annihilation (e.g. \citealt{ripamonti2007MNRAS.375.1399R}) that may reduce the gas content of small halos and influence our conclusion. However, it is beyond the scope of this paper to investigate the explicit influence at current time.

In this work, we only consider the influence of the small-scale power spectrum enhancement  on halo mass functions, ignoring the effects on the IGM. In principle, the enhancement also boosts the clumping of the IGM, so in C25 model the $\bar{n}_{\rm rec}$ in Eq.~(\ref{eq:excursion minihalo}) would be larger, consuming more ionizing photons. This will further strengthen  our conclusion.

Although we use the C25 model as demonstration, our methods are applicable for any power spectrum models with small-scale enhancement. Different scales and shapes of the enhancement may result in reionization histories and ionization field morphologies different from C25 model, but the conclusion \--- large-scale 21 cm signal statistics can reveal the enhancement, would always hold.

\section*{Acknowledgments}
We thank the anonymous referee very much for the constructive suggestions that help to improve our paper. This work is supported by the National SKA Program of China Nos. 2020SKA0110402 \& 2020SKA0110401, the NSFC International (Regional) Cooperation and Exchange Project No. 12361141814, the NSFC Key Program No. 12533002, NSFC grant No. 12421003, and the Project Supported by the Specialized Research Fund for State Key Laboratory of Radio Astronomy and Technology.
\bibliography{ref}{}

@ARTICLE{jenkins2001,
       author = {{Jenkins}, A. and {Frenk}, C.~S. and {White}, S.~D.~M. and {Colberg}, J.~M. and {Cole}, S. and {Evrard}, A.~E. and {Couchman}, H.~M.~P. and {Yoshida}, N.},
        title = "{The mass function of dark matter haloes}",
      journal = {\mnras},
         year = 2001,
        month = feb,
       volume = {321},
       number = {2},
        pages = {372-384},
          doi = {10.1046/j.1365-8711.2001.04029.x},
archivePrefix = {arXiv},
       eprint = {astro-ph/0005260},
 primaryClass = {astro-ph},
       adsurl = {https://ui.adsabs.harvard.edu/abs/2001MNRAS.321..372J}
}

@ARTICLE{st1999,
       author = {{Sheth}, Ravi K. and {Tormen}, Giuseppe},
        title = "{Large-scale bias and the peak background split}",
      journal = {\mnras},
         year = 1999,
        month = sep,
       volume = {308},
       number = {1},
        pages = {119-126},
          doi = {10.1046/j.1365-8711.1999.02692.x},
archivePrefix = {arXiv},
       eprint = {astro-ph/9901122},
 primaryClass = {astro-ph},
       adsurl = {https://ui.adsabs.harvard.edu/abs/1999MNRAS.308..119S}
}

@ARTICLE{Cielo2025,
       author = {{Cielo}, Mattia and {Mangano}, Gianpiero and {Pisanti}, Ofelia and {Wands}, David},
        title = "{Steepest growth in the primordial power spectrum from excited states at a sudden transition}",
      journal = {\jcap},
         year = 2025,
        month = apr,
       volume = {2025},
       number = {4},
          eid = {007},
        pages = {007},
          doi = {10.1088/1475-7516/2025/04/007},
archivePrefix = {arXiv},
       eprint = {2410.22154},
 primaryClass = {astro-ph.CO},
       adsurl = {https://ui.adsabs.harvard.edu/abs/2025JCAP...04..007C}
}

@ARTICLE{donnan2024,
       author = {{Donnan}, C.~T. and {McLure}, R.~J. and {Dunlop}, J.~S. and {McLeod}, D.~J. and {Magee}, D. and {Arellano-C{\'o}rdova}, K.~Z. and {Barrufet}, L. and {Begley}, R. and {Bowler}, R.~A.~A. and {Carnall}, A.~C. and {Cullen}, F. and {Ellis}, R.~S. and {Fontana}, A. and {Illingworth}, G.~D. and {Grogin}, N.~A. and {Hamadouche}, M.~L. and {Koekemoer}, A.~M. and {Liu}, F. -Y. and {Mason}, C. and {Santini}, P. and {Stanton}, T.~M.},
        title = "{JWST PRIMER: a new multifield determination of the evolving galaxy UV luminosity function at redshifts z ≃ 9 - 15}",
      journal = {\mnras},
         year = 2024,
        month = sep,
       volume = {533},
       number = {3},
        pages = {3222-3237},
          doi = {10.1093/mnras/stae2037},
archivePrefix = {arXiv},
       eprint = {2403.03171},
 primaryClass = {astro-ph.GA},
       adsurl = {https://ui.adsabs.harvard.edu/abs/2024MNRAS.533.3222D}
}

@ARTICLE{tacchella2018,
       author = {{Tacchella}, Sandro and {Bose}, Sownak and {Conroy}, Charlie and {Eisenstein}, Daniel J. and {Johnson}, Benjamin D.},
        title = "{A Redshift-independent Efficiency Model: Star Formation and Stellar Masses in Dark Matter Halos at z {\ensuremath{\gtrsim}} 4}",
      journal = {\apj},
         year = 2018,
        month = dec,
       volume = {868},
       number = {2},
          eid = {92},
        pages = {92},
          doi = {10.3847/1538-4357/aae8e0},
archivePrefix = {arXiv},
       eprint = {1806.03299},
 primaryClass = {astro-ph.GA},
       adsurl = {https://ui.adsabs.harvard.edu/abs/2018ApJ...868...92T}
}

@ARTICLE{mclure2013,
       author = {{McLure}, R.~J. and {Dunlop}, J.~S. and {Bowler}, R.~A.~A. and {Curtis-Lake}, E. and {Schenker}, M. and {Ellis}, R.~S. and {Robertson}, B.~E. and {Koekemoer}, A.~M. and {Rogers}, A.~B. and {Ono}, Y. and et al.},
        title = "{A new multifield determination of the galaxy luminosity function at z = 7-9 incorporating the 2012 Hubble Ultra-Deep Field imaging}",
      journal = {\mnras},
         year = 2013,
        month = jul,
       volume = {432},
       number = {4},
        pages = {2696-2716},
          doi = {10.1093/mnras/stt627},
archivePrefix = {arXiv},
       eprint = {1212.5222},
 primaryClass = {astro-ph.CO},
       adsurl = {https://ui.adsabs.harvard.edu/abs/2013MNRAS.432.2696M}
}

@ARTICLE{bowler2020,
       author = {{Bowler}, R.~A.~A. and {Jarvis}, M.~J. and {Dunlop}, J.~S. and {McLure}, R.~J. and {McLeod}, D.~J. and {Adams}, N.~J. and {Milvang-Jensen}, B. and {McCracken}, H.~J.},
        title = "{A lack of evolution in the very bright end of the galaxy luminosity function from z ≃ 8 to 10}",
      journal = {\mnras},
         year = 2020,
        month = apr,
       volume = {493},
       number = {2},
        pages = {2059-2084},
          doi = {10.1093/mnras/staa313},
archivePrefix = {arXiv},
       eprint = {1911.12832},
 primaryClass = {astro-ph.GA},
       adsurl = {https://ui.adsabs.harvard.edu/abs/2020MNRAS.493.2059B}
}

@ARTICLE{donnan2023,
       author = {{Donnan}, C.~T. and {McLeod}, D.~J. and {Dunlop}, J.~S. and {McLure}, R.~J. and {Carnall}, A.~C. and {Begley}, R. and {Cullen}, F. and {Hamadouche}, M.~L. and {Bowler}, R.~A.~A. and {Magee}, D. and et al.},
        title = "{The evolution of the galaxy UV luminosity function at redshifts z ≃ 8 - 15 from deep JWST and ground-based near-infrared imaging}",
      journal = {\mnras},
         year = 2023,
        month = feb,
       volume = {518},
       number = {4},
        pages = {6011-6040},
          doi = {10.1093/mnras/stac3472},
archivePrefix = {arXiv},
       eprint = {2207.12356},
 primaryClass = {astro-ph.GA},
       adsurl = {https://ui.adsabs.harvard.edu/abs/2023MNRAS.518.6011D}
}

@ARTICLE{mesinger2007,
       author = {{Mesinger}, Andrei and {Furlanetto}, Steven},
        title = "{Efficient Simulations of Early Structure Formation and Reionization}",
      journal = {\apj},
         year = 2007,
        month = nov,
       volume = {669},
       number = {2},
        pages = {663-675},
          doi = {10.1086/521806},
archivePrefix = {arXiv},
       eprint = {0704.0946},
 primaryClass = {astro-ph},
       adsurl = {https://ui.adsabs.harvard.edu/abs/2007ApJ...669..663M}
}

@ARTICLE{mesinger2011,
       author = {{Mesinger}, Andrei and {Furlanetto}, Steven and {Cen}, Renyue},
        title = "{21CMFAST: a fast, seminumerical simulation of the high-redshift 21-cm signal}",
      journal = {\mnras},
         year = 2011,
        month = feb,
       volume = {411},
       number = {2},
        pages = {955-972},
          doi = {10.1111/j.1365-2966.2010.17731.x},
archivePrefix = {arXiv},
       eprint = {1003.3878},
 primaryClass = {astro-ph.CO},
       adsurl = {https://ui.adsabs.harvard.edu/abs/2011MNRAS.411..955M}
}

@ARTICLE{park2019,
       author = {{Park}, Jaehong and {Mesinger}, Andrei and {Greig}, Bradley and {Gillet}, Nicolas},
        title = "{Inferring the astrophysics of reionization and cosmic dawn from galaxy luminosity functions and the 21-cm signal}",
      journal = {\mnras},
         year = 2019,
        month = mar,
       volume = {484},
       number = {1},
        pages = {933-949},
          doi = {10.1093/mnras/stz032},
archivePrefix = {arXiv},
       eprint = {1809.08995},
 primaryClass = {astro-ph.GA},
       adsurl = {https://ui.adsabs.harvard.edu/abs/2019MNRAS.484..933P}
}

@ARTICLE{fzh04,
       author = {{Furlanetto}, Steven R. and {Zaldarriaga}, Matias and {Hernquist}, Lars},
        title = "{The Growth of H II Regions During Reionization}",
      journal = {\apj},
         year = 2004,
        month = sep,
       volume = {613},
       number = {1},
        pages = {1-15},
          doi = {10.1086/423025},
archivePrefix = {arXiv},
       eprint = {astro-ph/0403697},
 primaryClass = {astro-ph},
       adsurl = {https://ui.adsabs.harvard.edu/abs/2004ApJ...613....1F}
}

@ARTICLE{shapiro2004,
       author = {{Shapiro}, Paul R. and {Iliev}, Ilian T. and {Raga}, Alejandro C.},
        title = "{Photoevaporation of cosmological minihaloes during reionization}",
      journal = {\mnras},
         year = 2004,
        month = mar,
       volume = {348},
       number = {3},
        pages = {753-782},
          doi = {10.1111/j.1365-2966.2004.07364.x},
archivePrefix = {arXiv},
       eprint = {astro-ph/0307266},
 primaryClass = {astro-ph},
       adsurl = {https://ui.adsabs.harvard.edu/abs/2004MNRAS.348..753S}
}

@ARTICLE{iliev2005,
       author = {{Iliev}, Ilian T. and {Shapiro}, Paul R. and {Raga}, Alejandro C.},
        title = "{Minihalo photoevaporation during cosmic reionization: evaporation times and photon consumption rates}",
      journal = {\mnras},
         year = 2005,
        month = aug,
       volume = {361},
       number = {2},
        pages = {405-414},
          doi = {10.1111/j.1365-2966.2005.09155.x},
archivePrefix = {arXiv},
       eprint = {astro-ph/0408408},
 primaryClass = {astro-ph},
       adsurl = {https://ui.adsabs.harvard.edu/abs/2005MNRAS.361..405I}
}

@ARTICLE{chan2024,
       author = {{Chan}, Tsang Keung and {Ben{\'\i}tez-Llambay}, Alejandro and {Theuns}, Tom and {Frenk}, Carlos and {Bower}, Richard},
        title = "{The impact and response of mini-haloes and the interhalo medium on cosmic reionization}",
      journal = {\mnras},
         year = 2024,
        month = feb,
       volume = {528},
       number = {2},
        pages = {1296-1326},
          doi = {10.1093/mnras/stae114},
archivePrefix = {arXiv},
       eprint = {2305.04959},
 primaryClass = {astro-ph.CO},
       adsurl = {https://ui.adsabs.harvard.edu/abs/2024MNRAS.528.1296C}
}

@ARTICLE{mo1996,
       author = {{Mo}, H.~J. and {White}, S.~D.~M.},
        title = "{An analytic model for the spatial clustering of dark matter haloes}",
      journal = {\mnras},
         year = 1996,
        month = sep,
       volume = {282},
       number = {2},
        pages = {347-361},
          doi = {10.1093/mnras/282.2.347},
archivePrefix = {arXiv},
       eprint = {astro-ph/9512127},
 primaryClass = {astro-ph},
       adsurl = {https://ui.adsabs.harvard.edu/abs/1996MNRAS.282..347M}
}

@ARTICLE{barkana2001,
       author = {{Barkana}, R. and {Loeb}, A.},
        title = "{In the beginning: the first sources of light and the reionization of the universe}",
      journal = {\physrep},
         year = 2001,
        month = jul,
       volume = {349},
       number = {2},
        pages = {125-238},
          doi = {10.1016/S0370-1573(01)00019-9},
archivePrefix = {arXiv},
       eprint = {astro-ph/0010468},
 primaryClass = {astro-ph},
       adsurl = {https://ui.adsabs.harvard.edu/abs/2001PhR...349..125B}
}

@ARTICLE{leitherer99,
       author = {{Leitherer}, Claus and {Schaerer}, Daniel and {Goldader}, Jeffrey D. and {Delgado}, Rosa M. Gonz{\'a}lez and {Robert}, Carmelle and {Kune}, Denis Foo and {de Mello}, Du{\'\i}lia F. and {Devost}, Daniel and {Heckman}, Timothy M.},
        title = "{Starburst99: Synthesis Models for Galaxies with Active Star Formation}",
      journal = {\apjs},
         year = 1999,
        month = jul,
       volume = {123},
       number = {1},
        pages = {3-40},
          doi = {10.1086/313233},
archivePrefix = {arXiv},
       eprint = {astro-ph/9902334},
 primaryClass = {astro-ph},
       adsurl = {https://ui.adsabs.harvard.edu/abs/1999ApJS..123....3L}
}

@ARTICLE{vzquez2005,
       author = {{V{\'a}zquez}, Gerardo A. and {Leitherer}, Claus},
        title = "{Optimization of Starburst99 for Intermediate-Age and Old Stellar Populations}",
      journal = {\apj},
         year = 2005,
        month = mar,
       volume = {621},
       number = {2},
        pages = {695-717},
          doi = {10.1086/427866},
archivePrefix = {arXiv},
       eprint = {astro-ph/0412491},
 primaryClass = {astro-ph},
       adsurl = {https://ui.adsabs.harvard.edu/abs/2005ApJ...621..695V}
}

@ARTICLE{leitherer2009,
       author = {{Leitherer}, Claus and {Chen}, Julia},
        title = "{Starburst99 for Windows}",
      journal = {\na},
         year = 2009,
        month = may,
       volume = {14},
       number = {4},
        pages = {356-362},
          doi = {10.1016/j.newast.2008.10.007},
archivePrefix = {arXiv},
       eprint = {0811.2396},
 primaryClass = {astro-ph},
       adsurl = {https://ui.adsabs.harvard.edu/abs/2009NewA...14..356L}
}

@ARTICLE{leitherer2014,
       author = {{Leitherer}, Claus and {Ekstr{\"o}m}, Sylvia and {Meynet}, Georges and {Schaerer}, Daniel and {Agienko}, Katerina B. and {Levesque}, Emily M.},
        title = "{The Effects of Stellar Rotation. II. A Comprehensive Set of Starburst99 Models}",
      journal = {\apjs},
         year = 2014,
        month = may,
       volume = {212},
       number = {1},
          eid = {14},
        pages = {14},
          doi = {10.1088/0067-0049/212/1/14},
archivePrefix = {arXiv},
       eprint = {1403.5444},
 primaryClass = {astro-ph.GA},
       adsurl = {https://ui.adsabs.harvard.edu/abs/2014ApJS..212...14L}
}

@ARTICLE{kroupa2001,
       author = {{Kroupa}, Pavel},
        title = "{On the variation of the initial mass function}",
      journal = {\mnras},
         year = 2001,
        month = apr,
       volume = {322},
       number = {2},
        pages = {231-246},
          doi = {10.1046/j.1365-8711.2001.04022.x},
archivePrefix = {arXiv},
       eprint = {astro-ph/0009005},
 primaryClass = {astro-ph},
       adsurl = {https://ui.adsabs.harvard.edu/abs/2001MNRAS.322..231K}
}

@ARTICLE{shull2012,
       author = {{Shull}, J. Michael and {Harness}, Anthony and {Trenti}, Michele and {Smith}, Britton D.},
        title = "{Critical Star Formation Rates for Reionization: Full Reionization Occurs at Redshift z {\ensuremath{\approx}} 7}",
      journal = {\apj},
         year = 2012,
        month = mar,
       volume = {747},
       number = {2},
          eid = {100},
        pages = {100},
          doi = {10.1088/0004-637X/747/2/100},
       adsurl = {https://ui.adsabs.harvard.edu/abs/2012ApJ...747..100S}
}

@ARTICLE{yb1970,
       author = {{Zel'dovich}, Ya. B.},
        title = "{Gravitational instability: An approximate theory for large density perturbations.}",
      journal = {\aap},
         year = 1970,
        month = mar,
       volume = {5},
        pages = {84-89},
       adsurl = {https://ui.adsabs.harvard.edu/abs/1970A&A.....5...84Z}
}

@ARTICLE{planck18,
       author = {{Planck Collaboration} and {Aghanim}, N. and {Akrami}, Y. and {Ashdown}, M. and {Aumont}, J. and {Baccigalupi}, C. and {Ballardini}, M. and {Banday}, A.~J. and {Barreiro}, R.~B. and {Bartolo}, N. and {Basak}, S. and {Battye}, R. and {Benabed}, K. and {Bernard}, J. -P. and {Bersanelli}, M. and {Bielewicz}, P. and {Bock}, J.~J. and {Bond}, J.~R. and {Borrill}, J. and {Bouchet}, F.~R. and {Boulanger}, F. and {Bucher}, M. and {Burigana}, C. and {Butler}, R.~C. and {Calabrese}, E. and {Cardoso}, J. -F. and {Carron}, J. and {Challinor}, A. and {Chiang}, H.~C. and {Chluba}, J. and {Colombo}, L.~P.~L. and {Combet}, C. and {Contreras}, D. and {Crill}, B.~P. and {Cuttaia}, F. and {de Bernardis}, P. and {de Zotti}, G. and {Delabrouille}, J. and {Delouis}, J. -M. and {Di Valentino}, E. and {Diego}, J.~M. and {Dor{\'e}}, O. and {Douspis}, M. and {Ducout}, A. and {Dupac}, X. and {Dusini}, S. and {Efstathiou}, G. and {Elsner}, F. and {En{\ss}lin}, T.~A. and {Eriksen}, H.~K. and {Fantaye}, Y. and {Farhang}, M. and {Fergusson}, J. and {Fernandez-Cobos}, R. and {Finelli}, F. and {Forastieri}, F. and {Frailis}, M. and {Fraisse}, A.~A. and {Franceschi}, E. and {Frolov}, A. and {Galeotta}, S. and {Galli}, S. and {Ganga}, K. and {G{\'e}nova-Santos}, R.~T. and {Gerbino}, M. and {Ghosh}, T. and {Gonz{\'a}lez-Nuevo}, J. and {G{\'o}rski}, K.~M. and {Gratton}, S. and {Gruppuso}, A. and {Gudmundsson}, J.~E. and {Hamann}, J. and {Handley}, W. and {Hansen}, F.~K. and {Herranz}, D. and {Hildebrandt}, S.~R. and {Hivon}, E. and {Huang}, Z. and {Jaffe}, A.~H. and {Jones}, W.~C. and {Karakci}, A. and {Keih{\"a}nen}, E. and {Keskitalo}, R. and {Kiiveri}, K. and {Kim}, J. and {Kisner}, T.~S. and {Knox}, L. and {Krachmalnicoff}, N. and {Kunz}, M. and {Kurki-Suonio}, H. and {Lagache}, G. and {Lamarre}, J. -M. and {Lasenby}, A. and {Lattanzi}, M. and {Lawrence}, C.~R. and {Le Jeune}, M. and {Lemos}, P. and {Lesgourgues}, J. and {Levrier}, F. and {Lewis}, A. and {Liguori}, M. and {Lilje}, P.~B. and {Lilley}, M. and {Lindholm}, V. and {L{\'o}pez-Caniego}, M. and {Lubin}, P.~M. and {Ma}, Y. -Z. and {Mac{\'\i}as-P{\'e}rez}, J.~F. and {Maggio}, G. and {Maino}, D. and {Mandolesi}, N. and {Mangilli}, A. and {Marcos-Caballero}, A. and {Maris}, M. and {Martin}, P.~G. and {Martinelli}, M. and {Mart{\'\i}nez-Gonz{\'a}lez}, E. and {Matarrese}, S. and {Mauri}, N. and {McEwen}, J.~D. and {Meinhold}, P.~R. and {Melchiorri}, A. and {Mennella}, A. and {Migliaccio}, M. and {Millea}, M. and {Mitra}, S. and {Miville-Desch{\^e}nes}, M. -A. and {Molinari}, D. and {Montier}, L. and {Morgante}, G. and {Moss}, A. and {Natoli}, P. and {N{\o}rgaard-Nielsen}, H.~U. and {Pagano}, L. and {Paoletti}, D. and {Partridge}, B. and {Patanchon}, G. and {Peiris}, H.~V. and {Perrotta}, F. and {Pettorino}, V. and {Piacentini}, F. and {Polastri}, L. and {Polenta}, G. and {Puget}, J. -L. and {Rachen}, J.~P. and {Reinecke}, M. and {Remazeilles}, M. and {Renzi}, A. and {Rocha}, G. and {Rosset}, C. and {Roudier}, G. and {Rubi{\~n}o-Mart{\'\i}n}, J.~A. and {Ruiz-Granados}, B. and {Salvati}, L. and {Sandri}, M. and {Savelainen}, M. and {Scott}, D. and {Shellard}, E.~P.~S. and {Sirignano}, C. and {Sirri}, G. and {Spencer}, L.~D. and {Sunyaev}, R. and {Suur-Uski}, A. -S. and {Tauber}, J.~A. and {Tavagnacco}, D. and {Tenti}, M. and {Toffolatti}, L. and {Tomasi}, M. and {Trombetti}, T. and {Valenziano}, L. and {Valiviita}, J. and {Van Tent}, B. and {Vibert}, L. and {Vielva}, P. and {Villa}, F. and {Vittorio}, N. and {Wandelt}, B.~D. and {Wehus}, I.~K. and {White}, M. and {White}, S.~D.~M. and {Zacchei}, A. and {Zonca}, A.},
        title = "{Planck 2018 results. VI. Cosmological parameters}",
      journal = {\aap},
         year = 2020,
        month = sep,
       volume = {641},
          eid = {A6},
        pages = {A6},
          doi = {10.1051/0004-6361/201833910},
archivePrefix = {arXiv},
       eprint = {1807.06209},
 primaryClass = {astro-ph.CO},
       adsurl = {https://ui.adsabs.harvard.edu/abs/2020A&A...641A...6P}
}

@ARTICLE{shull2010,
       author = {{Shull}, J. Michael and {France}, Kevin and {Danforth}, Charles W. and {Smith}, Britton and {Tumlinson}, Jason},
        title = "{HST/COS Observations of the Quasar HE 2347-4342: Probing the Epoch of He II Patchy Reionization at Redshifts z = 2.4-2.9}",
      journal = {\apj},
         year = 2010,
        month = oct,
       volume = {722},
       number = {2},
        pages = {1312-1324},
          doi = {10.1088/0004-637X/722/2/1312},
       adsurl = {https://ui.adsabs.harvard.edu/abs/2010ApJ...722.1312S}
}

@ARTICLE{jin2023,
       author = {{Jin}, Xiangyu and {Yang}, Jinyi and {Fan}, Xiaohui and {Wang}, Feige and {Ba{\~n}ados}, Eduardo and {Bian}, Fuyan and {Davies}, Frederick B. and {Eilers}, Anna-Christina and {Farina}, Emanuele Paolo and {Hennawi}, Joseph F. and {Pacucci}, Fabio and {Venemans}, Bram and {Walter}, Fabian},
        title = "{(Nearly) Model-independent Constraints on the Neutral Hydrogen Fraction in the Intergalactic Medium at z   5-7 Using Dark Pixel Fractions in Ly{\ensuremath{\alpha}} and Ly{\ensuremath{\beta}} Forests}",
      journal = {\apj},
         year = 2023,
        month = jan,
       volume = {942},
       number = {2},
          eid = {59},
        pages = {59},
          doi = {10.3847/1538-4357/aca678},
archivePrefix = {arXiv},
       eprint = {2211.12613},
 primaryClass = {astro-ph.CO},
       adsurl = {https://ui.adsabs.harvard.edu/abs/2023ApJ...942...59J}
}

@ARTICLE{umeda2025,
       author = {{Umeda}, Hiroya and {Ouchi}, Masami and {Kageura}, Yuta and {Harikane}, Yuichi and {Nakane}, Minami and {Thai}, Tran Thi and {Nakajima}, Kimihiko},
        title = "{Probing the Cosmic Reionization History with JWST: Gunn-Peterson and Ly$α$ Damping Wing Absorption at $4.5 < z < 13$}",
      journal = {arXiv e-prints},
         year = 2025,
        month = apr,
          eid = {arXiv:2504.04683},
        pages = {arXiv:2504.04683},
          doi = {10.48550/arXiv.2504.04683},
archivePrefix = {arXiv},
       eprint = {2504.04683},
 primaryClass = {astro-ph.GA},
       adsurl = {https://ui.adsabs.harvard.edu/abs/2025arXiv250404683U}
}

@ARTICLE{greig2024,
       author = {{Greig}, B. and {Mesinger}, A. and {Ba{\~n}ados}, E. and {Becker}, G.~D. and {Bosman}, S.~E.~I. and {Chen}, H. and {Davies}, F.~B. and {D'Odorico}, V. and {Eilers}, A. -C. and {Gallerani}, S. and et al.},
        title = "{IGM damping wing constraints on the tail end of reionization from the enlarged XQR-30 sample}",
      journal = {\mnras},
         year = 2024,
        month = may,
       volume = {530},
       number = {3},
        pages = {3208-3227},
          doi = {10.1093/mnras/stae1080},
archivePrefix = {arXiv},
       eprint = {2404.12585},
 primaryClass = {astro-ph.CO},
       adsurl = {https://ui.adsabs.harvard.edu/abs/2024MNRAS.530.3208G}
}

@ARTICLE{mason2018,
       author = {{Mason}, Charlotte A. and {Treu}, Tommaso and {Dijkstra}, Mark and {Mesinger}, Andrei and {Trenti}, Michele and {Pentericci}, Laura and {de Barros}, Stephane and {Vanzella}, Eros},
        title = "{The Universe Is Reionizing at z {\ensuremath{\sim}} 7: Bayesian Inference of the IGM Neutral Fraction Using Ly{\ensuremath{\alpha}} Emission from Galaxies}",
      journal = {\apj},
         year = 2018,
        month = mar,
       volume = {856},
       number = {1},
          eid = {2},
        pages = {2},
          doi = {10.3847/1538-4357/aab0a7},
archivePrefix = {arXiv},
       eprint = {1709.05356},
 primaryClass = {astro-ph.CO},
       adsurl = {https://ui.adsabs.harvard.edu/abs/2018ApJ...856....2M}
}

@ARTICLE{greig2022,
       author = {{Greig}, Bradley and {Mesinger}, Andrei and {Davies}, Frederick B. and {Wang}, Feige and {Yang}, Jinyi and {Hennawi}, Joseph F.},
        title = "{IGM damping wing constraints on reionization from covariance reconstruction of two z {\ensuremath{\gtrsim}} 7 QSOs}",
      journal = {\mnras},
         year = 2022,
        month = jun,
       volume = {512},
       number = {4},
        pages = {5390-5403},
          doi = {10.1093/mnras/stac825},
archivePrefix = {arXiv},
       eprint = {2112.04091},
 primaryClass = {astro-ph.CO},
       adsurl = {https://ui.adsabs.harvard.edu/abs/2022MNRAS.512.5390G}
}

@ARTICLE{hoag2019,
       author = {{Hoag}, A. and {Brada{\v{c}}}, M. and {Huang}, K. and {Mason}, C. and {Treu}, T. and {Schmidt}, K.~B. and {Trenti}, M. and {Strait}, V. and {Lemaux}, B.~C. and {Finney}, E.~Q. and et al.},
        title = "{Constraining the Neutral Fraction of Hydrogen in the IGM at Redshift 7.5}",
      journal = {\apj},
         year = 2019,
        month = jun,
       volume = {878},
       number = {1},
          eid = {12},
        pages = {12},
          doi = {10.3847/1538-4357/ab1de7},
archivePrefix = {arXiv},
       eprint = {1901.09001},
 primaryClass = {astro-ph.GA},
       adsurl = {https://ui.adsabs.harvard.edu/abs/2019ApJ...878...12H}
}

@ARTICLE{mason2019,
       author = {{Mason}, Charlotte A. and {Fontana}, Adriano and {Treu}, Tommaso and {Schmidt}, Kasper B. and {Hoag}, Austin and {Abramson}, Louis and {Amorin}, Ricardo and {Brada{\v{c}}}, Maru{\v{s}}a and {Guaita}, Lucia and {Jones}, Tucker and et al.},
        title = "{Inferences on the timeline of reionization at z {\ensuremath{\sim}} 8 from the KMOS Lens-Amplified Spectroscopic Survey}",
      journal = {\mnras},
         year = 2019,
        month = may,
       volume = {485},
       number = {3},
        pages = {3947-3969},
          doi = {10.1093/mnras/stz632},
archivePrefix = {arXiv},
       eprint = {1901.11045},
 primaryClass = {astro-ph.CO},
       adsurl = {https://ui.adsabs.harvard.edu/abs/2019MNRAS.485.3947M}
}

@ARTICLE{bruton2023,
       author = {{Bruton}, Sean and {Lin}, Yu-Heng and {Scarlata}, Claudia and {Hayes}, Matthew J.},
        title = "{The Universe is at Most 88\% Neutral at z = 10.6}",
      journal = {\apjl},
         year = 2023,
        month = jun,
       volume = {949},
       number = {2},
          eid = {L40},
        pages = {L40},
          doi = {10.3847/2041-8213/acd5d0},
archivePrefix = {arXiv},
       eprint = {2303.03419},
 primaryClass = {astro-ph.GA},
       adsurl = {https://ui.adsabs.harvard.edu/abs/2023ApJ...949L..40B}
}

@ARTICLE{curtis2023,
       author = {{Curtis-Lake}, Emma and {Carniani}, Stefano and {Cameron}, Alex and {Charlot}, Stephane and {Jakobsen}, Peter and {Maiolino}, Roberto and {Bunker}, Andrew and {Witstok}, Joris and {Smit}, Renske and {Chevallard}, Jacopo and et al.},
        title = "{Spectroscopic confirmation of four metal-poor galaxies at z = 10.3-13.2}",
      journal = {Nature Astronomy},
         year = 2023,
        month = may,
       volume = {7},
        pages = {622-632},
          doi = {10.1038/s41550-023-01918-w},
archivePrefix = {arXiv},
       eprint = {2212.04568},
 primaryClass = {astro-ph.GA},
       adsurl = {https://ui.adsabs.harvard.edu/abs/2023NatAs...7..622C}
}

@article{murray2024, doi = {10.21105/joss.06501}, url = {https://doi.org/10.21105/joss.06501}, year = {2024}, publisher = {The Open Journal}, volume = {9}, number = {97}, pages = {6501}, author = {Murray, Steven G. and Pober, Jonathan and Kolopanis, Matthew}, title = {21cmSense v2: A modular, open-source 21 cm sensitivity calculator}, journal = {Journal of Open Source Software} }

@ARTICLE{pober2013,
       author = {{Pober}, Jonathan C. and {Parsons}, Aaron R. and {DeBoer}, David R. and {McDonald}, Patrick and {McQuinn}, Matthew and {Aguirre}, James E. and {Ali}, Zaki and {Bradley}, Richard F. and {Chang}, Tzu-Ching and {Morales}, Miguel F.},
        title = "{The Baryon Acoustic Oscillation Broadband and Broad-beam Array: Design Overview and Sensitivity Forecasts}",
      journal = {\aj},
         year = 2013,
        month = mar,
       volume = {145},
       number = {3},
          eid = {65},
        pages = {65},
          doi = {10.1088/0004-6256/145/3/65},
archivePrefix = {arXiv},
       eprint = {1210.2413},
 primaryClass = {astro-ph.CO},
       adsurl = {https://ui.adsabs.harvard.edu/abs/2013AJ....145...65P}
}

@ARTICLE{pober2014,
       author = {{Pober}, Jonathan C. and {Liu}, Adrian and {Dillon}, Joshua S. and {Aguirre}, James E. and {Bowman}, Judd D. and {Bradley}, Richard F. and {Carilli}, Chris L. and {DeBoer}, David R. and {Hewitt}, Jacqueline N. and {Jacobs}, Daniel C. and et al.},
        title = "{What Next-generation 21 cm Power Spectrum Measurements can Teach us About the Epoch of Reionization}",
      journal = {\apj},
         year = 2014,
        month = feb,
       volume = {782},
       number = {2},
          eid = {66},
        pages = {66},
          doi = {10.1088/0004-637X/782/2/66},
archivePrefix = {arXiv},
       eprint = {1310.7031},
 primaryClass = {astro-ph.CO},
       adsurl = {https://ui.adsabs.harvard.edu/abs/2014ApJ...782...66P}
}

@ARTICLE{cruz2025,
       author = {{Cruz}, Hector Afonso G. and {Mu{\~n}oz}, Julian B. and {Sabti}, Nashwan and {Kamionkowski}, Marc},
        title = "{Effective model for the 21-cm signal with population III stars}",
      journal = {\prd},
         year = 2025,
        month = apr,
       volume = {111},
       number = {8},
          eid = {083503},
        pages = {083503},
          doi = {10.1103/PhysRevD.111.083503},
archivePrefix = {arXiv},
       eprint = {2407.18294},
 primaryClass = {astro-ph.CO},
       adsurl = {https://ui.adsabs.harvard.edu/abs/2025PhRvD.111h3503C}
}

@ARTICLE{giri2018,
       author = {{Giri}, Sambit K. and {Mellema}, Garrelt and {Dixon}, Keri L. and {Iliev}, Ilian T.},
        title = "{Bubble size statistics during reionization from 21-cm tomography}",
      journal = {\mnras},
         year = 2018,
        month = jan,
       volume = {473},
       number = {3},
        pages = {2949-2964},
          doi = {10.1093/mnras/stx2539},
archivePrefix = {arXiv},
       eprint = {1706.00665},
 primaryClass = {astro-ph.CO},
       adsurl = {https://ui.adsabs.harvard.edu/abs/2018MNRAS.473.2949G}
}

@ARTICLE{iliev2006,
       author = {{Iliev}, I.~T. and {Mellema}, G. and {Pen}, U.-L. and {Merz}, H. and {Shapiro}, P.~R. and {Alvarez}, M.~A.},
        title = "{Simulating cosmic reionization at large scales - I. The geometry of reionization}",
      journal = {\mnras},
         year = 2006,
        month = jul,
       volume = {369},
       number = {4},
        pages = {1625-1638},
          doi = {10.1111/j.1365-2966.2006.10502.x},
archivePrefix = {arXiv},
       eprint = {astro-ph/0512187},
 primaryClass = {astro-ph},
       adsurl = {https://ui.adsabs.harvard.edu/abs/2006MNRAS.369.1625I}
}

@ARTICLE{zahn2007,
       author = {{Zahn}, Oliver and {Lidz}, Adam and {McQuinn}, Matthew and {Dutta}, Suvendra and {Hernquist}, Lars and {Zaldarriaga}, Matias and {Furlanetto}, Steven R.},
        title = "{Simulations and Analytic Calculations of Bubble Growth during Hydrogen Reionization}",
      journal = {\apj},
         year = 2007,
        month = jan,
       volume = {654},
       number = {1},
        pages = {12-26},
          doi = {10.1086/509597},
archivePrefix = {arXiv},
       eprint = {astro-ph/0604177},
 primaryClass = {astro-ph},
       adsurl = {https://ui.adsabs.harvard.edu/abs/2007ApJ...654...12Z}
}

@ARTICLE{giri2020,
       author = {{Giri}, Sambit and {Mellema}, Garrelt and {Jensen}, Hannes},
        title = "{Tools21cm: A python package to analyse the large-scale 21-cm signal from the Epoch of Reionization and Cosmic Dawn}",
      journal = {The Journal of Open Source Software},
         year = 2020,
        month = aug,
       volume = {5},
       number = {52},
          eid = {2363},
        pages = {2363},
          doi = {10.21105/joss.02363},
       adsurl = {https://ui.adsabs.harvard.edu/abs/2020JOSS....5.2363G}
}

@ARTICLE{lewis2000,
       author = {{Lewis}, Antony and {Challinor}, Anthony and {Lasenby}, Anthony},
        title = "{Efficient Computation of Cosmic Microwave Background Anisotropies in Closed Friedmann-Robertson-Walker Models}",
      journal = {\apj},
         year = 2000,
        month = aug,
       volume = {538},
       number = {2},
        pages = {473-476},
          doi = {10.1086/309179},
archivePrefix = {arXiv},
       eprint = {astro-ph/9911177},
 primaryClass = {astro-ph},
       adsurl = {https://ui.adsabs.harvard.edu/abs/2000ApJ...538..473L}
}

@ARTICLE{bbks,
       author = {{Bardeen}, J.~M. and {Bond}, J.~R. and {Kaiser}, N. and {Szalay}, A.~S.},
        title = "{The Statistics of Peaks of Gaussian Random Fields}",
      journal = {\apj},
         year = 1986,
        month = may,
       volume = {304},
        pages = {15},
          doi = {10.1086/164143},
       adsurl = {https://ui.adsabs.harvard.edu/abs/1986ApJ...304...15B}
}

@ARTICLE{barkana2002,
       author = {{Barkana}, Rennan and {Loeb}, Abraham},
        title = "{Effective Screening Due to Minihalos during the Epoch of Reionization}",
      journal = {\apj},
         year = 2002,
        month = oct,
       volume = {578},
       number = {1},
        pages = {1-11},
          doi = {10.1086/342313},
archivePrefix = {arXiv},
       eprint = {astro-ph/0204139},
 primaryClass = {astro-ph},
       adsurl = {https://ui.adsabs.harvard.edu/abs/2002ApJ...578....1B}
}

@ARTICLE{ciardi2006,
       author = {{Ciardi}, B. and {Scannapieco}, E. and {Stoehr}, F. and {Ferrara}, A. and {Iliev}, I.~T. and {Shapiro}, P.~R.},
        title = "{The effect of minihaloes on cosmic reionization}",
      journal = {\mnras},
         year = 2006,
        month = feb,
       volume = {366},
       number = {2},
        pages = {689-696},
          doi = {10.1111/j.1365-2966.2005.09908.x},
archivePrefix = {arXiv},
       eprint = {astro-ph/0511623},
 primaryClass = {astro-ph},
       adsurl = {https://ui.adsabs.harvard.edu/abs/2006MNRAS.366..689C}
}

@ARTICLE{yb2009,
       author = {{Yue}, Bin and {Ciardi}, Benedetta and {Scannapieco}, Evan and {Chen}, Xuelei},
        title = "{The contribution of the IGM and minihaloes to the 21-cm signal of reionization}",
      journal = {\mnras},
         year = 2009,
        month = oct,
       volume = {398},
       number = {4},
        pages = {2122-2133},
          doi = {10.1111/j.1365-2966.2009.15261.x},
archivePrefix = {arXiv},
       eprint = {0906.3105},
 primaryClass = {astro-ph.CO},
       adsurl = {https://ui.adsabs.harvard.edu/abs/2009MNRAS.398.2122Y}
}

@ARTICLE{my2020,
       author = {{Mao}, Yi and {Koda}, Jun and {Shapiro}, Paul R. and {Iliev}, Ilian T. and {Mellema}, Garrelt and {Park}, Hyunbae and {Ahn}, Kyungjin and {Bianco}, Michele},
        title = "{The impact of inhomogeneous subgrid clumping on cosmic reionization}",
      journal = {\mnras},
         year = 2020,
        month = jan,
       volume = {491},
       number = {2},
        pages = {1600-1621},
          doi = {10.1093/mnras/stz2986},
archivePrefix = {arXiv},
       eprint = {1906.02476},
 primaryClass = {astro-ph.CO},
       adsurl = {https://ui.adsabs.harvard.edu/abs/2020MNRAS.491.1600M}
}

@ARTICLE{klessen2023,
       author = {{Klessen}, Ralf S. and {Glover}, Simon C.~O.},
        title = "{The First Stars: Formation, Properties, and Impact}",
      journal = {\araa},
         year = 2023,
        month = aug,
       volume = {61},
        pages = {65-130},
          doi = {10.1146/annurev-astro-071221-053453},
archivePrefix = {arXiv},
       eprint = {2303.12500},
 primaryClass = {astro-ph.CO},
       adsurl = {https://ui.adsabs.harvard.edu/abs/2023ARA&A..61...65K}
}

@ARTICLE{zm2025,
       author = {{Zhang}, Meng and {Yue}, Bin and {Xu}, Yidong and {Ferrara}, Andrea},
        title = "{The Formation of Direct Collapse Black Holes at Cosmic Dawn and 21 cm Global Spectrum}",
      journal = {\apj},
         year = 2025,
        month = may,
       volume = {984},
       number = {2},
          eid = {100},
        pages = {100},
          doi = {10.3847/1538-4357/adc730},
archivePrefix = {arXiv},
       eprint = {2503.22130},
 primaryClass = {astro-ph.GA},
       adsurl = {https://ui.adsabs.harvard.edu/abs/2025ApJ...984..100Z}
}

@ARTICLE{mcbride2009,
       author = {{McBride}, James and {Fakhouri}, Onsi and {Ma}, Chung-Pei},
        title = "{Mass accretion rates and histories of dark matter haloes}",
      journal = {\mnras},
         year = 2009,
        month = oct,
       volume = {398},
       number = {4},
        pages = {1858-1868},
          doi = {10.1111/j.1365-2966.2009.15329.x},
archivePrefix = {arXiv},
       eprint = {0902.3659},
 primaryClass = {astro-ph.CO},
       adsurl = {https://ui.adsabs.harvard.edu/abs/2009MNRAS.398.1858M}
}

@ARTICLE{koprowski2018,
       author = {{Koprowski}, M.~P. and {Coppin}, K.~E.~K. and {Geach}, J.~E. and {McLure}, R.~J. and {Almaini}, O. and {Blain}, A.~W. and {Bremer}, M. and {Bourne}, N. and {Chapman}, S.~C. and {Conselice}, C.~J. and et al.},
        title = "{A direct calibration of thtae IRX-{\ensuremath{\beta}} relation in Lyman-break Galaxies at z = 3-5}",
      journal = {\mnras},
         year = 2018,
        month = oct,
       volume = {479},
       number = {4},
        pages = {4355-4366},
          doi = {10.1093/mnras/sty1527},
archivePrefix = {arXiv},
       eprint = {1801.00791},
 primaryClass = {astro-ph.GA},
       adsurl = {https://ui.adsabs.harvard.edu/abs/2018MNRAS.479.4355K}
}

@ARTICLE{vogelsberger2019,
       author = {{Vogelsberger}, Mark and {McKinnon}, Ryan and {O'Neil}, Stephanie and {Marinacci}, Federico and {Torrey}, Paul and {Kannan}, Rahul},
        title = "{Dust in and around galaxies: dust in cluster environments and its impact on gas cooling}",
      journal = {\mnras},
         year = 2019,
        month = aug,
       volume = {487},
       number = {4},
        pages = {4870-4883},
          doi = {10.1093/mnras/stz1644},
archivePrefix = {arXiv},
       eprint = {1811.05477},
 primaryClass = {astro-ph.GA},
       adsurl = {https://ui.adsabs.harvard.edu/abs/2019MNRAS.487.4870V}
}

@ARTICLE{bouwens2021,
       author = {{Bouwens}, R.~J. and {Oesch}, P.~A. and {Stefanon}, M. and {Illingworth}, G. and {Labb{\'e}}, I. and {Reddy}, N. and {Atek}, H. and {Montes}, M. and {Naidu}, R. and {Nanayakkara}, T. and et al.},
        title = "{New Determinations of the UV Luminosity Functions from z   9 to 2 Show a Remarkable Consistency with Halo Growth and a Constant Star Formation Efficiency}",
      journal = {\aj},
         year = 2021,
        month = aug,
       volume = {162},
       number = {2},
          eid = {47},
        pages = {47},
          doi = {10.3847/1538-3881/abf83e},
archivePrefix = {arXiv},
       eprint = {2102.07775},
 primaryClass = {astro-ph.GA},
       adsurl = {https://ui.adsabs.harvard.edu/abs/2021AJ....162...47B}
}

@ARTICLE{bouwen2023,
       author = {{Bouwens}, Rychard J. and {Stefanon}, Mauro and {Brammer}, Gabriel and {Oesch}, Pascal A. and {Herard-Demanche}, Thomas and {Illingworth}, Garth D. and {Matthee}, Jorryt and {Naidu}, Rohan P. and {van Dokkum}, Pieter G. and {van Leeuwen}, Ivana F.},
        title = "{Evolution of the UV LF from z   15 to z   8 using new JWST NIRCam medium-band observations over the HUDF/XDF}",
      journal = {\mnras},
         year = 2023,
        month = jul,
       volume = {523},
       number = {1},
        pages = {1036-1055},
          doi = {10.1093/mnras/stad1145},
archivePrefix = {arXiv},
       eprint = {2211.02607},
 primaryClass = {astro-ph.GA},
       adsurl = {https://ui.adsabs.harvard.edu/abs/2023MNRAS.523.1036B}
}

@ARTICLE{harikane2023,
       author = {{Harikane}, Yuichi and {Ouchi}, Masami and {Oguri}, Masamune and {Ono}, Yoshiaki and {Nakajima}, Kimihiko and {Isobe}, Yuki and {Umeda}, Hiroya and {Mawatari}, Ken and {Zhang}, Yechi},
        title = "{A Comprehensive Study of Galaxies at z   9-16 Found in the Early JWST Data: Ultraviolet Luminosity Functions and Cosmic Star Formation History at the Pre-reionization Epoch}",
      journal = {\apjs},
         year = 2023,
        month = mar,
       volume = {265},
       number = {1},
          eid = {5},
        pages = {5},
          doi = {10.3847/1538-4365/acaaa9},
archivePrefix = {arXiv},
       eprint = {2208.01612},
 primaryClass = {astro-ph.GA},
       adsurl = {https://ui.adsabs.harvard.edu/abs/2023ApJS..265....5H}
}

@ARTICLE{finkelstein2015,
       author = {{Finkelstein}, Steven L. and {Ryan}, Jr., Russell E. and {Papovich}, Casey and {Dickinson}, Mark and {Song}, Mimi and {Somerville}, Rachel S. and {Ferguson}, Henry C. and {Salmon}, Brett and {Giavalisco}, Mauro and {Koekemoer}, Anton M. and et al.},
        title = "{The Evolution of the Galaxy Rest-frame Ultraviolet Luminosity Function over the First Two Billion Years}",
      journal = {\apj},
         year = 2015,
        month = sep,
       volume = {810},
       number = {1},
          eid = {71},
        pages = {71},
          doi = {10.1088/0004-637X/810/1/71},
archivePrefix = {arXiv},
       eprint = {1410.5439},
 primaryClass = {astro-ph.GA},
       adsurl = {https://ui.adsabs.harvard.edu/abs/2015ApJ...810...71F}
}

@ARTICLE{furlanetto2006,
       author = {{Furlanetto}, Steven R. and {Oh}, S. Peng and {Briggs}, Frank H.},
        title = "{Cosmology at low frequencies: The 21 cm transition and the high-redshift Universe}",
      journal = {\physrep},
         year = 2006,
        month = oct,
       volume = {433},
       number = {4-6},
        pages = {181-301},
          doi = {10.1016/j.physrep.2006.08.002},
archivePrefix = {arXiv},
       eprint = {astro-ph/0608032},
 primaryClass = {astro-ph},
       adsurl = {https://ui.adsabs.harvard.edu/abs/2006PhR...433..181F}
}

@ARTICLE{friedrich2011,
       author = {{Friedrich}, Martina M. and {Mellema}, Garrelt and {Alvarez}, Marcelo A. and {Shapiro}, Paul R. and {Iliev}, Ilian T.},
        title = "{Topology and sizes of H II regions during cosmic reionization}",
      journal = {\mnras},
         year = 2011,
        month = may,
       volume = {413},
       number = {2},
        pages = {1353-1372},
          doi = {10.1111/j.1365-2966.2011.18219.x},
archivePrefix = {arXiv},
       eprint = {1006.2016},
 primaryClass = {astro-ph.CO},
       adsurl = {https://ui.adsabs.harvard.edu/abs/2011MNRAS.413.1353F}
}

@ARTICLE{furlanetto2005,
       author = {{Furlanetto}, Steven R. and {Oh}, S. Peng},
        title = "{Taxing the rich: recombinations and bubble growth during reionization}",
      journal = {\mnras},
         year = 2005,
        month = nov,
       volume = {363},
       number = {3},
        pages = {1031-1048},
          doi = {10.1111/j.1365-2966.2005.09505.x},
archivePrefix = {arXiv},
       eprint = {astro-ph/0505065},
 primaryClass = {astro-ph},
       adsurl = {https://ui.adsabs.harvard.edu/abs/2005MNRAS.363.1031F}
}

@ARTICLE{naidu2022,
       author = {{Naidu}, Rohan P. and {Oesch}, Pascal A. and {van Dokkum}, Pieter and {Nelson}, Erica J. and {Suess}, Katherine A. and {Brammer}, Gabriel and {Whitaker}, Katherine E. and {Illingworth}, Garth and {Bouwens}, Rychard and {Tacchella}, Sandro and et al.},
        title = "{Two Remarkably Luminous Galaxy Candidates at z {\ensuremath{\approx}} 10-12 Revealed by JWST}",
      journal = {\apjl},
         year = 2022,
        month = nov,
       volume = {940},
       number = {1},
          eid = {L14},
        pages = {L14},
          doi = {10.3847/2041-8213/ac9b22},
archivePrefix = {arXiv},
       eprint = {2207.09434},
 primaryClass = {astro-ph.GA},
       adsurl = {https://ui.adsabs.harvard.edu/abs/2022ApJ...940L..14N}
}

@ARTICLE{castellano2022,
       author = {{Castellano}, Marco and {Fontana}, Adriano and {Treu}, Tommaso and {Santini}, Paola and {Merlin}, Emiliano and {Leethochawalit}, Nicha and {Trenti}, Michele and {Vanzella}, Eros and {Mestric}, Uros and {Bonchi}, Andrea and et al.},
        title = "{Early Results from GLASS-JWST. III. Galaxy Candidates at z  9-15}",
      journal = {\apjl},
         year = 2022,
        month = oct,
       volume = {938},
       number = {2},
          eid = {L15},
        pages = {L15},
          doi = {10.3847/2041-8213/ac94d0},
archivePrefix = {arXiv},
       eprint = {2207.09436},
 primaryClass = {astro-ph.GA},
       adsurl = {https://ui.adsabs.harvard.edu/abs/2022ApJ...938L..15C}
}

@ARTICLE{finkelstein2022,
       author = {{Finkelstein}, Steven L. and {Bagley}, Micaela B. and {Arrabal Haro}, Pablo and {Dickinson}, Mark and {Ferguson}, Henry C. and {Kartaltepe}, Jeyhan S. and {Papovich}, Casey and {Burgarella}, Denis and {Kocevski}, Dale D. and {Huertas-Company}, Marc and et al.},
        title = "{A Long Time Ago in a Galaxy Far, Far Away: A Candidate z {\ensuremath{\sim}} 12 Galaxy in Early JWST CEERS Imaging}",
      journal = {\apjl},
         year = 2022,
        month = dec,
       volume = {940},
       number = {2},
          eid = {L55},
        pages = {L55},
          doi = {10.3847/2041-8213/ac966e},
archivePrefix = {arXiv},
       eprint = {2207.12474},
 primaryClass = {astro-ph.GA},
       adsurl = {https://ui.adsabs.harvard.edu/abs/2022ApJ...940L..55F}
}

@ARTICLE{atek2023,
       author = {{Atek}, Hakim and {Shuntov}, Marko and {Furtak}, Lukas J. and {Richard}, Johan and {Kneib}, Jean-Paul and {Mahler}, Guillaume and {Zitrin}, Adi and {McCracken}, H.~J. and {Charlot}, St{\'e}phane and {Chevallard}, Jacopo and et al.},
        title = "{Revealing galaxy candidates out to z   16 with JWST observations of the lensing cluster SMACS0723}",
      journal = {\mnras},
         year = 2023,
        month = feb,
       volume = {519},
       number = {1},
        pages = {1201-1220},
          doi = {10.1093/mnras/stac3144},
archivePrefix = {arXiv},
       eprint = {2207.12338},
 primaryClass = {astro-ph.GA},
       adsurl = {https://ui.adsabs.harvard.edu/abs/2023MNRAS.519.1201A}
}

@ARTICLE{bradley2023,
       author = {{Bradley}, Larry D. and {Coe}, Dan and {Brammer}, Gabriel and {Furtak}, Lukas J. and {Larson}, Rebecca L. and {Kokorev}, Vasily and {Andrade-Santos}, Felipe and {Bhatawdekar}, Rachana and {Brada{\v{c}}}, Maru{\v{s}}a and {Broadhurst}, Tom and et al.},
        title = "{High-redshift Galaxy Candidates at z = 9-10 as Revealed by JWST Observations of WHL0137-08}",
      journal = {\apj},
         year = 2023,
        month = sep,
       volume = {955},
       number = {1},
          eid = {13},
        pages = {13},
          doi = {10.3847/1538-4357/acecfe},
archivePrefix = {arXiv},
       eprint = {2210.01777},
 primaryClass = {astro-ph.GA},
       adsurl = {https://ui.adsabs.harvard.edu/abs/2023ApJ...955...13B}
}

@ARTICLE{pritchard2012,
       author = {{Pritchard}, Jonathan R. and {Loeb}, Abraham},
        title = "{21 cm cosmology in the 21st century}",
      journal = {Reports on Progress in Physics},
         year = 2012,
        month = aug,
       volume = {75},
       number = {8},
          eid = {086901},
        pages = {086901},
          doi = {10.1088/0034-4885/75/8/086901},
archivePrefix = {arXiv},
       eprint = {1109.6012},
 primaryClass = {astro-ph.CO},
       adsurl = {https://ui.adsabs.harvard.edu/abs/2012RPPh...75h6901P}
}

@ARTICLE{Adamo2025NatAs,
       author = {{Adamo}, Angela and {Atek}, Hakim and {Bagley}, Micaela B. and {Ba{\~n}ados}, Eduardo and {Barrow}, Kirk S.~S. and {Berg}, Danielle A. and {Bezanson}, Rachel and {Brada{\v{c}}}, Maru{\v{s}}a and {Brammer}, Gabriel and {Carnall}, Adam C. and {Chisholm}, John and {Coe}, Dan and {Dayal}, Pratika and {Eisenstein}, Daniel J. and {Eldridge}, Jan J. and {Ferrara}, Andrea and {Fujimoto}, Seiji and {Graaff}, Anna de and {Habouzit}, Melanie and {Hutchison}, Taylor A. and {Kartaltepe}, Jeyhan S. and {Kassin}, Susan A. and {Kriek}, Mariska and {Labb{\'e}}, Ivo and {Maiolino}, Roberto and {Marques-Chaves}, Rui and {Maseda}, Michael V. and {Mason}, Charlotte and {Matthee}, Jorryt and {McQuinn}, Kristen B.~W. and {Meynet}, Georges and {Naidu}, Rohan P. and {Oesch}, Pascal A. and {Pentericci}, Laura and {P{\'e}rez-Gonz{\'a}lez}, Pablo G. and {Rigby}, Jane R. and {Roberts-Borsani}, Guido and {Schaerer}, Daniel and {Shapley}, Alice E. and {Stark}, Daniel P. and {Stiavelli}, Massimo and {Strom}, Allison L. and {Vanzella}, Eros and {Wang}, Feige and {Wilkins}, Stephen M. and {Williams}, Christina C. and {Willott}, Chris J. and {Wylezalek}, Dominika and {Nota}, Antonella},
        title = "{The first billion years according to JWST}",
      journal = {Nature Astronomy},
         year = 2025,
        month = aug,
       volume = {9},
        pages = {1134-1147},
          doi = {10.1038/s41550-025-02624-5},
archivePrefix = {arXiv},
       eprint = {2405.21054},
 primaryClass = {astro-ph.GA},
       adsurl = {https://ui.adsabs.harvard.edu/abs/2025NatAs...9.1134A}
}

@ARTICLE{yue2015,
       author = {{Yue}, B. and {Ferrara}, A. and {Pallottini}, A. and {Gallerani}, S. and {Vallini}, L.},
        title = "{Intensity mapping of [C II] emission from early galaxies}",
      journal = {\mnras},
         year = 2015,
        month = jul,
       volume = {450},
       number = {4},
        pages = {3829-3839},
          doi = {10.1093/mnras/stv933},
archivePrefix = {arXiv},
       eprint = {1504.06530},
 primaryClass = {astro-ph.GA},
       adsurl = {https://ui.adsabs.harvard.edu/abs/2015MNRAS.450.3829Y}
}

@ARTICLE{gsm08,
       author = {{de Oliveira-Costa}, Ang{\'e}lica and {Tegmark}, Max and {Gaensler}, B.~M. and {Jonas}, Justin and {Landecker}, T.~L. and {Reich}, Patricia},
        title = "{A model of diffuse Galactic radio emission from 10 MHz to 100 GHz}",
      journal = {\mnras},
         year = 2008,
        month = jul,
       volume = {388},
       number = {1},
        pages = {247-260},
          doi = {10.1111/j.1365-2966.2008.13376.x},
archivePrefix = {arXiv},
       eprint = {0802.1525},
 primaryClass = {astro-ph},
       adsurl = {https://ui.adsabs.harvard.edu/abs/2008MNRAS.388..247D}
}

@ARTICLE{gsm16,
       author = {{Zheng}, H. and {Tegmark}, M. and {Dillon}, J.~S. and {Kim}, D.~A. and {Liu}, A. and {Neben}, A.~R. and {Jonas}, J. and {Reich}, P. and {Reich}, W.},
        title = "{An improved model of diffuse galactic radio emission from 10 MHz to 5 THz}",
      journal = {\mnras},
         year = 2017,
        month = jan,
       volume = {464},
       number = {3},
        pages = {3486-3497},
          doi = {10.1093/mnras/stw2525},
archivePrefix = {arXiv},
       eprint = {1605.04920},
 primaryClass = {astro-ph.CO},
       adsurl = {https://ui.adsabs.harvard.edu/abs/2017MNRAS.464.3486Z}
}

@ARTICLE{Wang2020Natur,
       author = {{Wang}, J. and {Bose}, S. and {Frenk}, C.~S. and {Gao}, L. and {Jenkins}, A. and {Springel}, V. and {White}, S.~D.~M.},
        title = "{Universal structure of dark matter haloes over a mass range of 20 orders of magnitude}",
      journal = {\nat},
         year = 2020,
        month = sep,
       volume = {585},
       number = {7823},
        pages = {39-42},
          doi = {10.1038/s41586-020-2642-9},
archivePrefix = {arXiv},
       eprint = {1911.09720},
 primaryClass = {astro-ph.CO},
       adsurl = {https://ui.adsabs.harvard.edu/abs/2020Natur.585...39W}
}

@ARTICLE{Diemand2005Natur,
       author = {{Diemand}, J. and {Moore}, B. and {Stadel}, J.},
        title = "{Earth-mass dark-matter haloes as the first structures in the early Universe}",
      journal = {\nat},
         year = 2005,
        month = jan,
       volume = {433},
       number = {7024},
        pages = {389-391},
          doi = {10.1038/nature03270},
archivePrefix = {arXiv},
       eprint = {astro-ph/0501589},
 primaryClass = {astro-ph},
       adsurl = {https://ui.adsabs.harvard.edu/abs/2005Natur.433..389D}
}

@ARTICLE{viel2005PhRvD,
       author = {{Viel}, Matteo and {Lesgourgues}, Julien and {Haehnelt}, Martin G. and {Matarrese}, Sabino and {Riotto}, Antonio},
        title = "{Constraining warm dark matter candidates including sterile neutrinos and light gravitinos with WMAP and the Lyman-{\ensuremath{\alpha}} forest}",
      journal = {\prd},
         year = 2005,
        month = mar,
       volume = {71},
       number = {6},
          eid = {063534},
        pages = {063534},
          doi = {10.1103/PhysRevD.71.063534},
archivePrefix = {arXiv},
       eprint = {astro-ph/0501562},
 primaryClass = {astro-ph},
       adsurl = {https://ui.adsabs.harvard.edu/abs/2005PhRvD..71f3534V}
}

@ARTICLE{Barkana2001ApJ,
       author = {{Barkana}, Rennan and {Haiman}, Zolt{\'a}n and {Ostriker}, Jeremiah P.},
        title = "{Constraints on Warm Dark Matter from Cosmological Reionization}",
      journal = {\apj},
         year = 2001,
        month = sep,
       volume = {558},
       number = {2},
        pages = {482-496},
          doi = {10.1086/322393},
archivePrefix = {arXiv},
       eprint = {astro-ph/0102304},
 primaryClass = {astro-ph},
       adsurl = {https://ui.adsabs.harvard.edu/abs/2001ApJ...558..482B}
}

@ARTICLE{Yue2012ApJ,
       author = {{Yue}, Bin and {Chen}, Xuelei},
        title = "{Reionization in the Warm Dark Matter Model}",
      journal = {\apj},
         year = 2012,
        month = mar,
       volume = {747},
       number = {2},
          eid = {127},
        pages = {127},
          doi = {10.1088/0004-637X/747/2/127},
archivePrefix = {arXiv},
       eprint = {1201.3686},
 primaryClass = {astro-ph.CO},
       adsurl = {https://ui.adsabs.harvard.edu/abs/2012ApJ...747..127Y}
}

@ARTICLE{Bode2001ApJ,
       author = {{Bode}, Paul and {Ostriker}, Jeremiah P. and {Turok}, Neil},
        title = "{Halo Formation in Warm Dark Matter Models}",
      journal = {\apj},
         year = 2001,
        month = jul,
       volume = {556},
       number = {1},
        pages = {93-107},
          doi = {10.1086/321541},
archivePrefix = {arXiv},
       eprint = {astro-ph/0010389},
 primaryClass = {astro-ph},
       adsurl = {https://ui.adsabs.harvard.edu/abs/2001ApJ...556...93B}
}

@ARTICLE{Smith2011PhRvD,
       author = {{Smith}, Robert E. and {Markovic}, Katarina},
        title = "{Testing the warm dark matter paradigm with large-scale structures}",
      journal = {\prd},
         year = 2011,
        month = sep,
       volume = {84},
       number = {6},
          eid = {063507},
        pages = {063507},
          doi = {10.1103/PhysRevD.84.063507},
archivePrefix = {arXiv},
       eprint = {1103.2134},
 primaryClass = {astro-ph.CO},
       adsurl = {https://ui.adsabs.harvard.edu/abs/2011PhRvD..84f3507S}
}

@ARTICLE{Gong2023ApJ,
       author = {{Gong}, Yan and {Yue}, Bin and {Cao}, Ye and {Chen}, Xuelei},
        title = "{Fuzzy Dark Matter as a Solution to Reconcile the Stellar Mass Density of High-z Massive Galaxies and Reionization History}",
      journal = {\apj},
         year = 2023,
        month = apr,
       volume = {947},
       number = {1},
          eid = {28},
        pages = {28},
          doi = {10.3847/1538-4357/acc109},
archivePrefix = {arXiv},
       eprint = {2209.13757},
 primaryClass = {astro-ph.CO},
       adsurl = {https://ui.adsabs.harvard.edu/abs/2023ApJ...947...28G}
}

@ARTICLE{Liu2025arXiv,
       author = {{Liu}, Shihang and {Liu}, Yilin and {Peng}, Bowen and {Xie}, Mengzhou and {Liu}, Zelong and {Li}, Bohua and {Mao}, Yi},
        title = "{Constraining fuzzy dark matter with the 21-cm power spectrum from Cosmic Dawn and Reionization}",
      journal = {arXiv e-prints},
         year = 2025,
        month = aug,
          eid = {arXiv:2508.10176},
        pages = {arXiv:2508.10176},
          doi = {10.48550/arXiv.2508.10176},
archivePrefix = {arXiv},
       eprint = {2508.10176},
 primaryClass = {astro-ph.CO},
       adsurl = {https://ui.adsabs.harvard.edu/abs/2025arXiv250810176L}
}

@ARTICLE{zhangxin2023,
       author = {{Zhang}, Xin and {Yue}, Bin and {Shi}, Yuan and {Wu}, Fengquan and {Chen}, Xuelei},
        title = "{On Measuring the 21 cm Global Spectrum of the Cosmic Dawn with an Interferometer Array}",
      journal = {\apj},
         year = 2023,
        month = mar,
       volume = {945},
       number = {2},
          eid = {109},
        pages = {109},
          doi = {10.3847/1538-4357/acb6fe},
archivePrefix = {arXiv},
       eprint = {2301.12223},
 primaryClass = {astro-ph.CO},
       adsurl = {https://ui.adsabs.harvard.edu/abs/2023ApJ...945..109Z}
}

@ARTICLE{Jones2021ApJ,
       author = {{Jones}, Dana and {Palatnick}, Skyler and {Chen}, Richard and {Beane}, Angus and {Lidz}, Adam},
        title = "{Fuzzy Dark Matter and the 21 cm Power Spectrum}",
      journal = {\apj},
         year = 2021,
        month = may,
       volume = {913},
       number = {1},
          eid = {7},
        pages = {7},
          doi = {10.3847/1538-4357/abf0a9},
archivePrefix = {arXiv},
       eprint = {2101.07177},
 primaryClass = {astro-ph.CO},
       adsurl = {https://ui.adsabs.harvard.edu/abs/2021ApJ...913....7J}
}

@ARTICLE{Nadler2025ApJ,
       author = {{Nadler}, Ethan O. and {Gluscevic}, Vera and {Benson}, Andrew},
        title = "{The Effects of Linear Matter Power Spectrum Enhancement on Dark Matter Substructure}",
      journal = {\apj},
         year = 2025,
        month = nov,
       volume = {993},
       number = {1},
          eid = {17},
        pages = {17},
          doi = {10.3847/1538-4357/ae073b},
archivePrefix = {arXiv},
       eprint = {2507.16889},
 primaryClass = {astro-ph.CO},
       adsurl = {https://ui.adsabs.harvard.edu/abs/2025ApJ...993...17N}
}

@ARTICLE{Inman2023PhRvD,
       author = {{Inman}, Derek and {Kohri}, Kazunori},
        title = "{Enhanced small-scale structure in the cosmic dark ages}",
      journal = {\prd},
         year = 2023,
        month = jun,
       volume = {107},
       number = {12},
          eid = {123513},
        pages = {123513},
          doi = {10.1103/PhysRevD.107.123513},
archivePrefix = {arXiv},
       eprint = {2207.14735},
 primaryClass = {astro-ph.CO},
       adsurl = {https://ui.adsabs.harvard.edu/abs/2023PhRvD.107l3513I}
}

@ARTICLE{Endsley2020MNRAS,
       author = {{Endsley}, Ryan and {Behroozi}, Peter and {Stark}, Daniel P. and {Williams}, Christina C. and {Robertson}, Brant E. and {Rieke}, Marcia and {Gottl{\"o}ber}, Stefan and {Yepes}, Gustavo},
        title = "{Clustering with JWST: Constraining galaxy host halo masses, satellite quenching efficiencies, and merger rates at z = 4-10}",
      journal = {\mnras},
         year = 2020,
        month = mar,
       volume = {493},
       number = {1},
        pages = {1178-1196},
          doi = {10.1093/mnras/staa324},
archivePrefix = {arXiv},
       eprint = {1907.02546},
 primaryClass = {astro-ph.GA},
       adsurl = {https://ui.adsabs.harvard.edu/abs/2020MNRAS.493.1178E}
}

@ARTICLE{Chabanier2019MNRAS,
       author = {{Chabanier}, Sol{\`e}ne and {Millea}, Marius and {Palanque-Delabrouille}, Nathalie},
        title = "{Matter power spectrum: from Ly {\ensuremath{\alpha}} forest to CMB scales}",
      journal = {\mnras},
         year = 2019,
        month = oct,
       volume = {489},
       number = {2},
        pages = {2247-2253},
          doi = {10.1093/mnras/stz2310},
archivePrefix = {arXiv},
       eprint = {1905.08103},
 primaryClass = {astro-ph.CO},
       adsurl = {https://ui.adsabs.harvard.edu/abs/2019MNRAS.489.2247C}
}

@ARTICLE{Viel2013PhRvD,
       author = {{Viel}, Matteo and {Becker}, George D. and {Bolton}, James S. and {Haehnelt}, Martin G.},
        title = "{Warm dark matter as a solution to the small scale crisis: New constraints from high redshift Lyman-{\ensuremath{\alpha}} forest data}",
      journal = {\prd},
         year = 2013,
        month = aug,
       volume = {88},
       number = {4},
          eid = {043502},
        pages = {043502},
          doi = {10.1103/PhysRevD.88.043502},
archivePrefix = {arXiv},
       eprint = {1306.2314},
 primaryClass = {astro-ph.CO},
       adsurl = {https://ui.adsabs.harvard.edu/abs/2013PhRvD..88d3502V}
}

@ARTICLE{Villasenor2023PhRvD,
       author = {{Villasenor}, Bruno and {Robertson}, Brant and {Madau}, Piero and {Schneider}, Evan},
        title = "{New constraints on warm dark matter from the Lyman-{\ensuremath{\alpha}} forest power spectrum}",
      journal = {\prd},
         year = 2023,
        month = jul,
       volume = {108},
       number = {2},
          eid = {023502},
        pages = {023502},
          doi = {10.1103/PhysRevD.108.023502},
archivePrefix = {arXiv},
       eprint = {2209.14220},
 primaryClass = {astro-ph.CO},
       adsurl = {https://ui.adsabs.harvard.edu/abs/2023PhRvD.108b3502V}
}

@ARTICLE{Irvic2017PhRvD,
       author = {{Ir{\v{s}}i{\v{c}}}, Vid and {Viel}, Matteo and {Haehnelt}, Martin G. and {Bolton}, James S. and {Cristiani}, Stefano and {Becker}, George D. and {D'Odorico}, Valentina and {Cupani}, Guido and {Kim}, Tae-Sun and {Berg}, Trystyn A.~M. and {L{\'o}pez}, Sebastian and {Ellison}, Sara and {Christensen}, Lise and {Denney}, Kelly D. and {Worseck}, G{\'a}bor},
        title = "{New constraints on the free-streaming of warm dark matter from intermediate and small scale Lyman-{\ensuremath{\alpha}} forest data}",
      journal = {\prd},
         year = 2017,
        month = jul,
       volume = {96},
       number = {2},
          eid = {023522},
        pages = {023522},
          doi = {10.1103/PhysRevD.96.023522},
archivePrefix = {arXiv},
       eprint = {1702.01764},
 primaryClass = {astro-ph.CO},
       adsurl = {https://ui.adsabs.harvard.edu/abs/2017PhRvD..96b3522I}
}

@ARTICLE{Sipple2025MNRAS,
       author = {{Sipple}, Jackson and {Lidz}, Adam and {Grin}, Daniel and {Sun}, Guochao},
        title = "{Fuzzy dark matter constraints from the Hubble Frontier Fields}",
      journal = {\mnras},
         year = 2025,
        month = apr,
       volume = {538},
       number = {3},
        pages = {1830-1842},
          doi = {10.1093/mnras/staf340},
archivePrefix = {arXiv},
       eprint = {2407.17059},
 primaryClass = {astro-ph.CO},
       adsurl = {https://ui.adsabs.harvard.edu/abs/2025MNRAS.538.1830S}
}

@ARTICLE{Lazare2024PhRvD,
       author = {{Lazare}, Hovav and {Flitter}, Jordan and {Kovetz}, Ely D.},
        title = "{Constraints on the fuzzy dark matter mass window from high-redshift observables}",
      journal = {\prd},
         year = 2024,
        month = dec,
       volume = {110},
       number = {12},
          eid = {123532},
        pages = {123532},
          doi = {10.1103/PhysRevD.110.123532},
archivePrefix = {arXiv},
       eprint = {2407.19549},
 primaryClass = {astro-ph.CO},
       adsurl = {https://ui.adsabs.harvard.edu/abs/2024PhRvD.110l3532L}
}

@ARTICLE{Rudakovskyi2021MNRAS,
       author = {{Rudakovskyi}, Anton and {Mesinger}, Andrei and {Savchenko}, Denys and {Gillet}, Nicolas},
        title = "{Constraints on warm dark matter from UV luminosity functions of high-z galaxies with Bayesian model comparison}",
      journal = {\mnras},
         year = 2021,
        month = oct,
       volume = {507},
       number = {2},
        pages = {3046-3056},
          doi = {10.1093/mnras/stab2333},
archivePrefix = {arXiv},
       eprint = {2104.04481},
 primaryClass = {astro-ph.CO},
       adsurl = {https://ui.adsabs.harvard.edu/abs/2021MNRAS.507.3046R}
}

@article{Allen1985,
    author = "Allen, Bruce",
    title = "{Vacuum States in de Sitter Space}",
    reportNumber = "UCSB-TH-3-1985",
    doi = "10.1103/PhysRevD.32.3136",
    journal = {\prd},
    volume = "32",
    pages = "3136",
    year = "1985"
}
\bibliographystyle{aasjournalv7}
 
\end{document}